\documentclass[aps,twocolumn,amsmath,amssymb,floatfix,pra,reprint,showpacs,footinbib,superscriptaddress,longbibliography]{revtex4-1}
\usepackage{amssymb}
\usepackage{amsmath}
\usepackage{dcolumn}
\usepackage{graphicx}
\usepackage{mathrsfs}
\usepackage{subfigure}
\usepackage{booktabs}
\usepackage{color}
\usepackage{docmute}
\usepackage{hyphenat}

\newcommand{\average}[1]{\mbox{$\langle#1\rangle$}}

\usepackage{url}
\usepackage[colorlinks]{hyperref}
\hypersetup{%
    plainpages=true,
    breaklinks=true,
    hypertexnames=false,
    pageanchor=true,
    bookmarksnumbered=true,
    colorlinks=true,
    linkcolor={blue},
    citecolor={red},
    urlcolor={blue},
    anchorcolor={black}
}

\begin{document}

\title{Emission of photon pairs by mechanical stimulation of the squeezed vacuum}

\author{Wei Qin}
\affiliation{Theoretical Quantum Physics Laboratory, RIKEN Cluster
for Pioneering Research, Wako-shi, Saitama 351-0198, Japan}

\author{Vincenzo Macr\`{i}}
\affiliation{Theoretical Quantum Physics Laboratory, RIKEN Cluster
for Pioneering Research, Wako-shi, Saitama 351-0198, Japan}

\author{Adam Miranowicz}
\affiliation{Theoretical Quantum Physics Laboratory, RIKEN Cluster
for Pioneering Research, Wako-shi, Saitama 351-0198, Japan}
\affiliation{Faculty of Physics, Adam Mickiewicz University,
61-614 Pozna\'n, Poland}

\author{Salvatore Savasta}
\affiliation{Theoretical Quantum Physics Laboratory, RIKEN Cluster
for Pioneering Research, Wako-shi, Saitama 351-0198, Japan}
\affiliation{Dipartimento di Scienze Matematiche e Informatiche,
Scienze Fisiche e Scienze della Terra, \\ Universit\`{a} di
Messina, I-98166 Messina, Italy}

\author{Franco Nori}
\affiliation{Theoretical Quantum Physics Laboratory, RIKEN Cluster
for Pioneering Research, Wako-shi, Saitama 351-0198, Japan}
\affiliation{Department of Physics, The University of Michigan,
Ann Arbor, Michigan 48109-1040, USA}

\begin{abstract}
To observe the dynamical Casimir effect (DCE) induced by a moving
mirror is a long-standing challenge because the mirror velocity
needs to approach the speed of light. Here, we present an
experimentally feasible method for observing this mechanical DCE
in an optomechanical system. It employs a detuned, parametric
driving to squeeze a cavity mode, so that the mechanical mode,
with a typical resonance frequency, can parametrically and
resonantly couple to the squeezed cavity mode, thus leading to a
resonantly amplified DCE in the squeezed frame. The DCE process
can be interpreted as {\it mechanically-induced two-photon
hyper-Raman scattering} in the laboratory frame. Specifically,
{\it a photon pair} of the parametric driving absorbs a single
phonon and then is scattered into an anti-Stokes sideband. We also
find that the squeezing, which additionally induces and amplifies
the DCE, can be extremely small. Our method requires neither an
ultra-high mechanical-oscillation frequency (i.e., a mirror moving
at nearly the speed of light) nor an ultrastrong single-photon
optomechanical coupling and, thus, could be implemented in a wide
range of physical systems.
\end{abstract}

\maketitle
\section{Introduction}

One of the most astonishing phenomena of nature, predicted by
quantum field theory, is that the quantum vacuum is not empty but
teems with virtual particles. Under certain conditions, these
vacuum fluctuations could be converted into real particles by
dynamical amplification mechanisms such as the Schwinger
process~\cite{schwinger1951gauge}, Hawking
radiation~\cite{hawking1974black}, and Unruh
effect~\cite{unruh1976notes}. The dynamical Casimir effect (DCE)
describes the creation of photons out of the quantum vacuum due to
a moving mirror~\cite{moore1970quantum, fulling1976radiation}. The
physics underlying the DCE is that the electromagnetic field
cannot adiabatically adapt to the time-dependent boundary
condition imposed by the mechanical motion of the mirror, such
that it occurs a mismatch of vacuum modes in time. This gives rise
to the emission of photon pairs from the vacuum and, at the same
time, to the equal-energy dissipation of the mechanical phonons.
Thus, according to energy conservation, the DCE can also be
understood as the energy conversion of the mechanical motion to
the electromagnetic field.

In order to detect the DCE, the mirror velocity is, however,
required to be close to the speed of
light~\cite{dodonov2010current,nation2012colloquium}. This
requirement is the main obstacle in observing the DCE. This
problem led to many alternative proposals, which replaced the
mechanical motion with an effective motion provided by, e.g.,
modulating dielectric properties of semiconductors or
superconductors~\cite{yablonovitch1989accelerating,lozovik1995parametric,crocce2004model,braggio2005novel,segev2007prospects},
modulating the ultrastrong light-matter coupling in cavity quantum
electrodynamics
(QED)~\cite{ciuti2005quantum,de2007quantum,de2009extracavity,garziano2013switching,hagenmuller2016all,de2017virtual,cirio2017amplified,e2018microscopic,kockum2019ultrastrong,forn2019ultrastrong},
or driving an optical parametric
oscillator~\cite{dezael2010analogue}. In particular, two
remarkable experimental verifications have recently been
implemented utilizing a superconducting quantum interference
device~\cite{nation2012colloquium,johansson2009dynamical,johansson2010dynamical,wilson2011observation,dalvit2011quantum,johansson2013nonclassical}
and a Josephson metamaterial~\cite{lahteenmaki2013dynamical},
respectively, to produce the effective motion. Despite such
achievements, implementing the DCE with a massive mechanical
mirror is still highly desirable for a more fundamental
understanding of the DCE physics. This is because the parametric
conversion of mechanical energy to photons, which is a key feature
of the DCE predicted in its original
proposals~\cite{moore1970quantum,
fulling1976radiation,dodonov2010current,nation2012colloquium}, can
be demonstrated in this case, contrary to proposals based on the
effective motion. However, owing to the serious problem mentioned
above (i.e., very fast oscillating mirror), such a radiation has
not yet been observed experimentally, although the DCE has been
predicted for almost fifty years. Here, we propose a novel
approach to this outstanding problem, and we show that in a
squeezed optomechanical system, a mirror oscillating at a common
frequency can induce an observable DCE.

The DCE can, in principle, also be directly implemented in
cavity-optomechanical
systems~\cite{lambrecht1996motion,dodonov1996generation,plunien2000dynamical,schaller2002dynamical,kim2006detectability,
de2013influence,macri2018nonperturbative,Sanz2018electromechanical,wang2018mechanically,settineri2019conversion}.
But it requires a mechanical frequency $\omega_{m}$ to be very
close to the cavity frequency $\omega_{c}$, or even a
single-photon optomechanical coupling $g_{0}$ to reach the
ultrastrong-coupling regime
$g_{0}/\omega_{m}\!\gtrsim\!0.1$~\cite{macri2018nonperturbative,settineri2019conversion}.
For typical parameters, $\omega_{m}\!\!\sim\!\!$~MHz is much
smaller than $\omega_{c}\!\!\sim\!\!$~THz ($\sim$~\!\!GHz) for
optical (microwave) cavities, and at the same time, achieving the
ultrastrong coupling is, currently, also a very challenging task
in optomechanical experiments. However, as we describe in this
manuscript, when squeezing the cavity~\cite{scully1997book}, the
squeezed-cavity-mode (SCM) frequency is tunable, such that the SCM
can parametrically and resonantly couple to a mechanical mode with
a typically available $\omega_{m}$. This enables an observable DCE
in the squeezed frame. Such a {\it mechanical DCE corresponds to
two-photon hyper-Raman scattering} in the laboratory frame.
Compared to one-photon Raman scattering typically demonstrated in
cavity optomechanics, this hyper-Raman scattering process
describes {\it a photon pair scattered into a higher energy mode
by absorbing a mechanical phonon}.

As opposed to previous mechanical-DCE proposals, our approach
requires {\it neither} an ultra-high mechanical frequency {\it
nor} an ultrastrong coupling. In addition, the model discussed
here is a generic optomechanical setup. Hence, with current
technologies our proposal could be realized in various physical
architectures, e.g., superconducting
resonators~\cite{xiang2013hybrid,gu2017microwave} and optical
cavities~\cite{reiserer2015cavity}. Furthermore, our proposal also
shows mechanically-induced two-photon hyper-Raman scattering,
which, to our knowledge, has not been considered before in cavity
optomechanics.

\begin{figure}[t]
\centering
\includegraphics[width=8.3cm,angle=0]{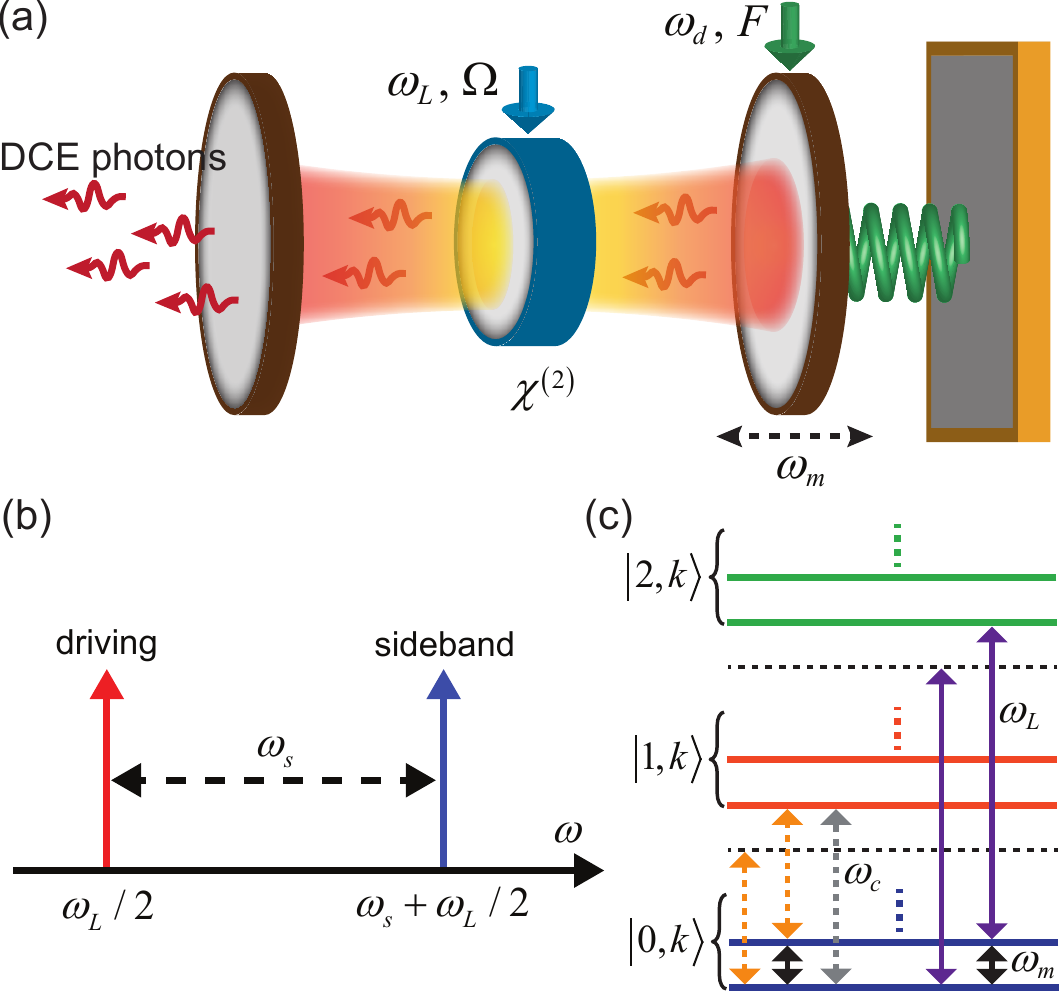}
\caption{(a) Setup for observing the mechanical dynamical Casimir
effect. In this optomechanical system, a $\chi^{\left(2\right)}$
nonlinear crystal driven at a frequency $\omega_{L}$ and amplitude
$\Omega$ is used to squeeze the cavity mode of frequency
$\omega_{c}$, and the mechanical resonator is driven by a force of
frequency $\omega_{d}$ and amplitude $F$. The DCE occurs in the
squeezed frame, and a large number of DCE photons, emitted from
the cavity, can be observed in the laboratory frame. (b)
Frequency-domain illustration of mechanically-induced two-photon
hyper-Raman scattering. The left arrow is the two-photon driving
($\omega_{L}/2$), and the right arrow is the squeezing-induced
anti-Stokes sideband ($\omega_{s}+\omega_{L}/2$). The horizontal
axis corresponds to the resonance frequency ($\omega$) and the
dashed double arrow to the detuning ($\omega_{s}$). (c) Level
diagram of the bare optomechanical system. The solid (dashed)
arrows indicate two-photon hyper Raman (one-photon Raman)
scattering processes induced by the optomechanical coupling. The
first number in the ket refers to the photon number and the second
to the phonon number $k$. We assume, for simplicity, that
$\Omega\ll\Delta$, such that the resonance condition is
$2\omega_{c}\approx\omega_{L}+\omega_{m}$.} \label{fig-schematic}
\end{figure}

\section{Model}

We consider an optomechanical system, as schematically depicted in
Fig.~\ref{fig-schematic}(a). The basic idea underlying our
proposal is to use a detuned two-photon driving, e.g., of
frequency $\omega_{L}$ and amplitude $\Omega$, to squeeze the
cavity mode. The driving results in parametric down conversion of
mechanical phonons to correlated cavity-photon pairs, which
corresponds to the DCE. Furthermore, the SCM frequency completely
depends on the detuning $\Delta=\omega_{c}-\omega_{L}/2$ and the
amplitude $\Omega$. This can be exploited to tune the parametric
phonon-photon coupling into resonance, determining a strong
amplification of the DCE. When the mechanical mode is driven,
e.g., at frequency $\omega_{d}$ and amplitude $F$, a strong
steady-state output-photon flux that is induced by the DCE can be
achieved.

To be specific, we consider the Hamiltonian
\begin{equation}
H=H_{\rm OM}+H_{\rm CD}+H_{\rm MD}.
\end{equation}
Here,
\begin{equation}
    H_{\rm OM}=\omega_{m}b^{\dag}b-g_{0}a^{\dag}a\left(b+b^{\dag}\right)
\end{equation}
describes a standard optomechanical coupling,
\begin{equation}
H_{\rm CD}=\Delta
a^{\dag}a+\frac{1}{2}\Omega\left(a^{2}+a^{\dag2}\right)
\end{equation}
a detuned two-photon cavity driving, and
\begin{equation}
H_{\rm
MD}=\frac{1}{2}F\left[\exp\left(i\omega_{d}t\right)b+\exp\left(-i\omega_{d}t\right)b^{\dag}\right]
\end{equation}
a single-phonon mechanical driving. The bare cavity mode $a$, when
parametrically driven, is squeezed with a squeezing parameter
\begin{equation}
r=\frac{1}{4}\ln\left(\frac{\Delta+\Omega}{\Delta-\Omega}\right)
\end{equation}
and accordingly, is transformed to a squeezed mode $a_{s}$, via
the Bogoliubov transformation~\cite{scully1997book}
\begin{equation}
a_{s}=\cosh\left(r\right)a+\sinh\left(r\right)a^{\dag}.
\end{equation}
Similar methods have been used for enhancing light-matter
interactions in cavity
optomechanics~\cite{lu2015squeezed,lemonde2016enhanced} and cavity
QED~\cite{qin2018exponentially,
        leroux2018enhancing}, but involving markedly different physical processes.
As a result, $H_{\rm CD}$ is diagonalized to $H_{\rm
CD}=\omega_{s}a_{s}^{\dag}a_{s}$, where
$\omega_{s}=\sqrt{\Delta^{2}-\Omega^{2}}$ is a controllable SCM
frequency. The optomechanical-coupling Hamiltonian is transformed,
in terms of $a_{s}$, to
\begin{equation}
    H_{\rm
        OM}=\left[-g_{\rm OM}a_{s}^{\dag}a_{s}+g_{\rm
        DCE}\left(a_{s}^{2}+a^{\dag2}_{s}\right)\right]\left(b+b^{\dag}\right),
\end{equation}
where $g_{\rm OM}=g_{0}\cosh\left(2r\right)$ is an effective
single-photon optomechanical coupling, and $g_{\rm
DCE}=g_{0}\sinh\left(2r\right)/2$ is a coupling associated with
the DCE. The dynamics under $H_{\rm OM}$ describes a mechanical
modulation of the boundary condition of the squeezed
field~\cite{law1995interaction,macri2018nonperturbative,di2017interaction}.
Under the rotating-wave approximation, the coherent dynamics of
the system is governed by an effective Hamiltonian,
\begin{align}\label{sq:effectiveH}
H_{\rm eff}=\;&\Delta_{s}a_{s}^{\dag}a_{s}+\Delta_{m}b^{\dag}b\nonumber\\
&+g_{\rm DCE}\left(a_{s}^{2}b^{\dag}+{\rm
H.c.}\right)+\frac{1}{2}F\left(b+b^{\dag}\right),
\end{align}
where $\Delta_{s}=\omega_{s}-\omega_{d}/2$ and
$\Delta_{m}=\omega_{m}-\omega_{d}$. We find that when
$\omega_{m}=2\omega_{s}$, the resonant DCE can be demonstrated,
and that the parametric energy conversion of the  mechanical
motion to the electromagnetic field, which was predicted in the
original DCE proposals, can therefore be observed. We also find
that the energy of emitted photons in the squeezed frame
completely originates from the mechanical motion. Thus,
parametrically driving the cavity without a moving
mirror~\cite{dezael2010analogue}, corresponding to $F=0$, {\it
cannot} excite the $a_{s}$ mode and {\it cannot} result in such a
parametric energy conversion from mechanics to light.

\section{Mechanically-induced two-photon hyper-Raman scattering}

More interestingly, the DCE in the squeezed frame can be
interpreted, in the laboratory frame, as mechanically-induced
two-photon hyper-Raman scattering. This hyper-Raman scattering is
an anti-Stokes process, as illustrated in
Fig.~\ref{fig-schematic}(b). According to the Bogoliubov
transformation, the squeezing gives rise to an anti-Stokes
sideband at frequency $\omega_{s}+\omega_{L}/2$ [right arrow in
Fig.~\ref{fig-schematic}(b)]. The two-photon driving at frequency
$\omega_{L}$ produces photon pairs at frequency $\omega_{L}/2$
[left arrow in Fig.~\ref{fig-schematic}(b)]. When mechanical
phonons at frequency $\omega_{m}=2\omega_{s}$ are present, a
driving photon pair is
    scattered into the anti-Stokes sideband, while simultaneously absorbing a phonon in the mechanical resonator. Because of their different frequency from the driving photon pairs, the anti-Stokes scattered photon pairs, which are referred to as the DCE photons, can be spectrally filtered from the driving photons, which are referred to as the noise photons.

    In cavity optomechanics, most of the experimental and theoretical studies are carried out under detuned {\it one-photon} driving of a cavity, so that the cavity field can be split into an average coherent amplitude and a fluctuating term. For a red-detuned driving, a driving photon can be scattered into the cavity resonance by absorbing a phonon. This process is viewed as {\it mechanically-induced one-photon Raman scattering} [dashed arrows in Fig.~\ref{fig-schematic}(c)]. As described above, our proposal instead exploits a red-detuned {\it two-photon} driving, and {\it the mechanical motion can induce two-photon hyper-Raman scattering}. In order to compare the two scattering processes more explicitly, we consider the limit  $\Omega\ll\Delta$. In this limit, the $a_{s}$ mode can be
approximated by the $a$ mode, i.e., $a_{s}\approx a$, and as a
result, the anti-Stokes sideband becomes the cavity resonance.
Correspondingly, the effective Hamiltonian $H_{\rm eff}$ becomes
\begin{align}\label{eq:widetilde_Heff}
\widetilde{H}_{\rm eff}=\;&\Delta_s a^{\dag}a+\omega_{m}b^{\dag}b\nonumber\\
&+g_{\rm DCE}\left(a^{2}b^{\dag}+{\rm
H.c.}\right)+\frac{1}{2}F\left(b+b^{\dag}\right).
\end{align}
Under the resonant condition $\omega_{m}=2\omega_{s}$ (i.e.,
2$\omega_{c}\approx\omega_{L}+\omega_{m}$), the dynamics described
by $\widetilde{H}_{\rm eff}$ shows that a driving photon pair,
rather than a single photon, is scattered into the cavity
resonance by absorbing a phonon [solid arrows in
Fig.~\ref{fig-schematic}(c)].

\begin{figure}[t]
    \centering
    \includegraphics[width=8.3cm]{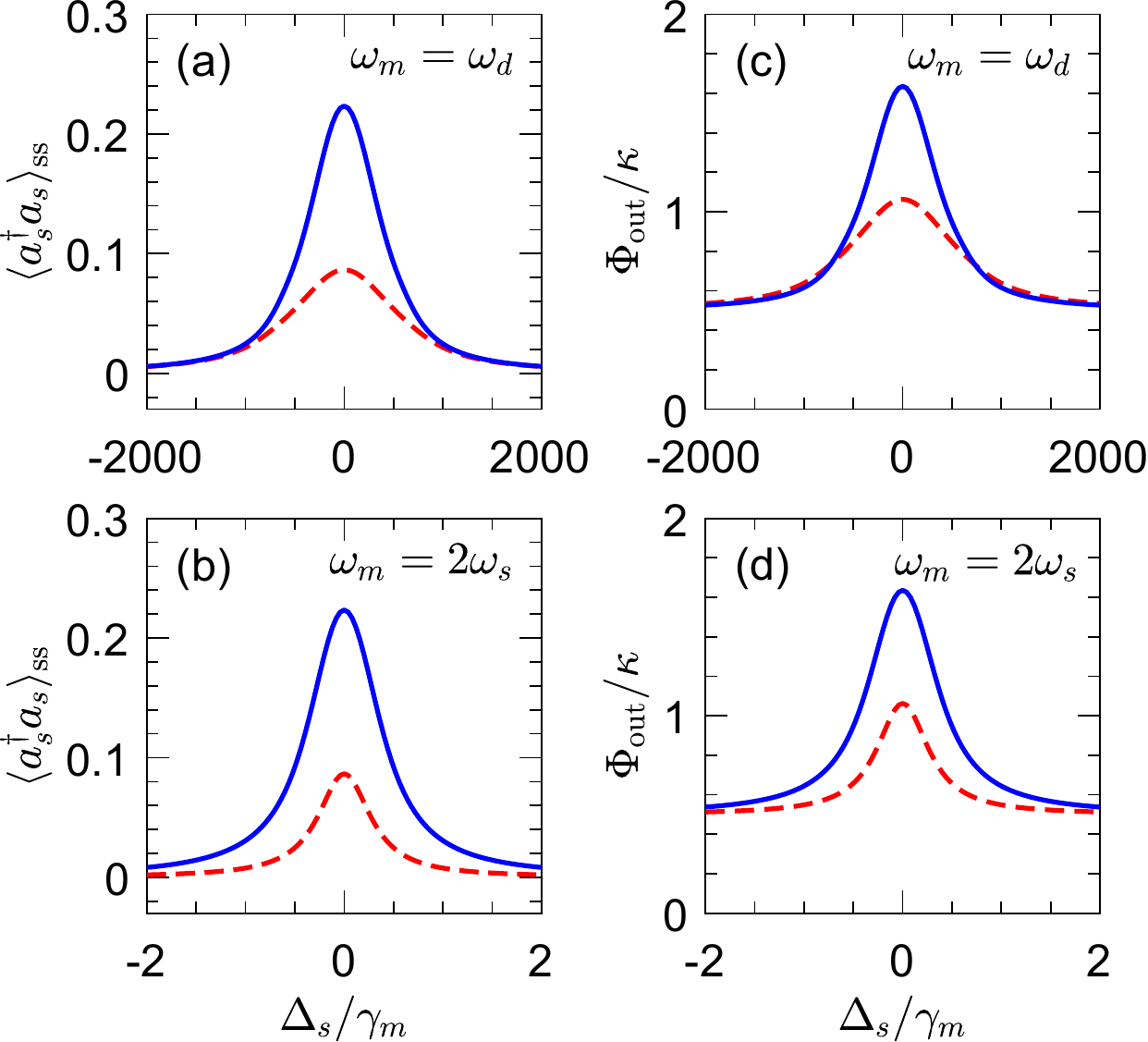}
    \caption{(a)-(b) Photon number
        $\langle a_{s}^{\dag}a_{s}\rangle_{\rm ss}$ and (c)-(d)
        photon flux $\Phi_{\rm out}$ versus detuning $\Delta_{s}$. The emergence of the
        resonance peaks indicates the occurrence of the dynamical Casimir effect. We
        assumed $\omega_{m}=\omega_{d}$ in (a) and (c), and
        $\omega_{m}=2\omega_{s}$ in (b) and (d). Solid curves correspond to $\kappa=500\gamma_{m}$ and dashed curves to $\kappa=1000\gamma_{m}$. In all plots, we assumed $g_{0}=10\gamma_{m}$, $F=15\gamma_{m}$,
        $\omega_{m}=10^{4}\gamma_{m}$, and
        $\sinh^{2}\left(r\right)=0.5$.}\label{fig-excitation-spectrum}
\end{figure}

\section{How to observe the dynamical Casimir effect}

In our approach, we squeeze the $a$ mode to make the effective
cavity frequency very close to the mechanical frequency. However,
this squeezing also inputs thermal noise and two-photon
correlation noise into the cavity. Although these undesired
effects are negligible in the weak-squeezing case (see below),
they can be completely eliminated by coupling a squeezed-vacuum
bath, e.g., with a squeezing parameter $r_{e}$ and a reference
phase $\theta_{e}$, to the $a$
mode~\cite{murch2013reduction,bartkowiak2014quantum,clark2017sideband,zeytinouglu2017engineering,vahlbruch2018laser}.
We assume that $r_{e}=r$ and $\theta_{e}=\pm n\pi$
($n=1,3,5,\cdots$), so that the $a_{s}$ mode is equivalently
coupled to a vacuum bath (see Appendix~\ref{sec:Optomechanical
master equation}). The full dynamics is therefore determined by
the standard master equation
\begin{equation}\label{eq:master-equation}
\dot{\rho}\left(t\right)=i\left[\rho\left(t\right),H_{\rm
eff}\right]-\frac{\kappa}{2}\mathcal{L}\left(a_{s}\right)\rho\left(t\right)-\frac{\gamma_{m}}{2}\mathcal{L}\left(b\right)\rho\left(t\right),
\end{equation}
where $\kappa$ and $\gamma_{m}$ are the cavity and mechanical loss
rates, respectively, and we have defined
\begin{equation}
\mathcal{L}\left(o\right)\rho\left(t\right)=o^{\dag}o\rho\left(t\right)-2o\rho\left(t\right)o^{\dag}+\rho\left(t\right)o^{\dag}o.
\end{equation}
We have also assumed that the mechanical resonator is coupled to a
zero-temperature bath (see Appendix~\ref{sec:dynamical Casimir
effect in the weak mechanical driving regime} for an analytical
discussion at finite temperatures). The SCM excitation spectrum
$\langle a_{s}^{\dag}a_{s}\rangle_{\rm
ss}\left(\Delta_{s}\right)$, where $\average{o}_{\rm ss}$
represents a steady-state average value, is plotted in
Figs.~\ref{fig-excitation-spectrum}(a) and
\ref{fig-excitation-spectrum}(b). Eliminating the
squeezing-induced noise ensures a zero background noise for the
excitation spectrum. If the mechanical resonator is driven, then
photons are excited from the vacuum, and according to energy
conservation, are emitted from the mechanical resonator, together
with a resonance peak in the excitation spectrum.

We now return to the original laboratory frame and consider the
steady-state output-photon flux. Because of the squeezing, the
steady-state intracavity photon number, $\average{a^{\dag}a}_{\rm
ss}$, in the laboratory frame includes two physical contributions,
i.e.,
\begin{equation}\label{eq:cavity_photon_number}
\average{a^{\dag}a}_{\rm ss}=\Phi_{\rm BGN}+\Phi_{\rm DCE},
\end{equation}
where $\Phi_{\rm BGN}=\sinh^{2}\left(r\right)$ is the number of
background-noise photons contained in the squeezed vacuum, and
\begin{equation}\label{eq:DCE_signal}
\Phi_{\rm DCE}=\average{a_{s}^{\dag}a_{s}}_{\rm
    ss}\cosh\left(2r\right)-{\rm Re}\left[\average{a_{s}^{2}}_{\rm
    ss}\right]\sinh\left(2r\right)
\end{equation}
is the number of DCE-induced photons. The output-photon flux is
then given by
\begin{equation}\label{eq:output_flux}
\Phi_{\rm out}=\kappa\left(\Phi_{\rm BGN}+\Phi_{\rm DCE}\right),
\end{equation}
according to the input-output relation. We plot the flux spectrum
$\Phi_{\rm out}\left(\Delta_{s}\right)$ in
Figs.~\ref{fig-excitation-spectrum}(c) and
\ref{fig-excitation-spectrum}(d). There exists a nonzero
background noise in the photon flux spectrum, as discussed
previously. Nevertheless, when driving the mechanical resonator,
the DCE-induced photons are emitted from the cavity, and a
resolved resonance peak can be observed. We find that the behavior
of the flux spectrum directly reflects that of the excitation
spectrum. Hence, the emergence of the resonance peak in the flux
spectrum can be considered as an experimentally observable
signature of the DCE.

\begin{figure}[t]
    \centering
    \includegraphics[width=8.0cm]{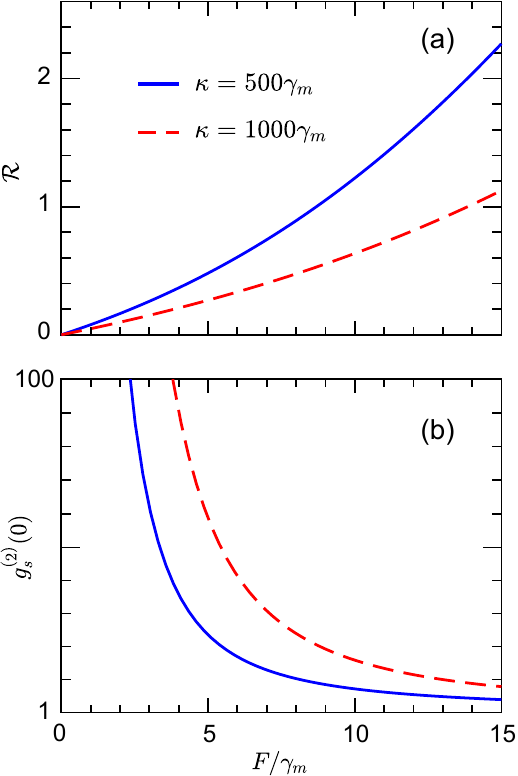}
    \caption{(a) Signal-to-noise ratio $\mathcal{R}$ and (b) equal-time correlation $g_{s}^{\left(2\right)}\!\left(0\right)$ as a function of the
        mechanical driving $F$ for $\kappa=500\gamma_{m}$ and
        $1000\gamma_{m}$. Note that the signal can still be resolved even if $\mathcal{R}<1$ with
        standard spectroscopic techniques used, e.g., for Raman signals. The $g_{s}^{\left(2\right)}\!\left(0\right)>1$ correlation implies that in the
        squeezed frame, the photons are created in pairs from the quantum
        vacuum. For both plots, we assumed that $g_{0}=10\gamma_{m}$,
        $\omega_{m}=\omega_{d}=2\omega_{s}$, and
        $\sinh^{2}\left(r\right)=0.5$.}\label{fig-signal-to-noise-ratio}
\end{figure}

Owing to the existence of the background noise in the flux
$\Phi_{\rm out}$, we now discuss the ability to resolve the DCE
signal $\Phi_{\rm DCE}$ from the background noise $\Phi_{\rm BGN}$
at resonance $\omega_{m}=\omega_{d}=2\omega_{s}$. In order to
quantify this, we typically employ the signal-to-noise ratio,
defined as
\begin{equation}\label{eq:signal_to_noise_ratio}
\mathcal{R}=\frac{\Phi_{\rm DCE}}{\Phi_{\rm BGN}}.
\end{equation}
The signal-resolved regime often requires $\mathcal{R}>1$,
allowing for a resolved DCE-signal detection. We find that, by
increasing the mechanical driving $F$, the signal $\Phi_{\rm DCE}$
becomes stronger, but at the same time, the noise $\Phi_{\rm BGN}$
remains unchanged. This enables an improvement in the
signal-to-noise ratio with the mechanical force. Consequently, the
desired signal can be directly driven from the unresolved to
resolved regime, as shown in
Fig.~\ref{fig-signal-to-noise-ratio}(a). Assuming a realistic
parameter $g_{0}=10\gamma_{m}$, we find that a mechanical driving
of $F=15\gamma_{m}$ is able to keep the ratio $\mathcal{R}$ above
$1$ for $\kappa\leq1000\gamma_{m}$. With these parameters, we can
obtain $\average{a_{s}^{\dag}a_{s}}_{\rm ss}\approx0.2$, as given
in Fig.~\ref{fig-excitation-spectrum}. Therefore, in the
laboratory frame, a cavity having a typical linewidth of
$\kappa/2\pi=2.0$~MHz could emit $\approx1.4\times10^{7}$ photons
per second, which is larger than the background photon emission
$\approx6.3\times10^{6}$ per second. The ratio $\mathcal{R}$ can
be made $\gg1$ as long as the driving $F$ is further increased, so
that the background noise can be even neglected compared to the
DCE signal. This is demonstrated in
Appendix~\ref{sec:Semi-classical treatment for the strong
mechanical driving}, where we make a semi-classical approximation
for investigating the DCE under a strong-$F$ drive. For
\begin{equation}
F\gg\left(g_{\rm DCE}+\kappa\gamma_{m}/4g_{\rm DCE}\right),
\end{equation}
the system behaves classically~\cite{wilson2010photon,
butera2019mechanical}, and quantum effects are negligible. Thus in
order to observe the DCE, such a regime needs to be avoided. Note,
however, that the signal can still be resolved even for
$\mathcal{R}<1$, if standard techniques of Raman spectroscopy are
used. This is because the background noise is due to driving
photons at frequency $\omega_{L}/2$, while the DCE photons have a
frequency $\omega_{s}+\omega_{L}/2$. The monotonic increase of the
flux $\Phi_{\rm out}$ at resonance with the driving $F$ can,
therefore, be considered as another signature of the mechanical
DCE in experiments.

The DCE photons are emitted in pairs, and could exhibit photon
bunching~\cite{johansson2010dynamical,macri2018nonperturbative,stassi2013spontaneous}.
The essential parameter characterizing this property is the
equal-time second-order correlation function,
\begin{equation}\label{eq:g2_correlation}
g^{\left(2\right)}_{s}\!\left(0\right)=\frac{\average{a_{s}^{\dag
2}a_{s}^{2}}_{\rm ss}}{\average{a_{s}^{\dag}a_{s}}_{\rm ss}^{2}}.
\end{equation}
We plot it as a function of the mechanical driving in
Fig.~\ref{fig-signal-to-noise-ratio}(b). We find that
\begin{equation}
g_{s}^{\left(2\right)}\!\left(0\right)\approx\frac{1}{2\average{a_{s}^{\dag}a_{s}}_{\rm
ss}}
\end{equation}
in the $F\rightarrow0$ limit, and $\approx1$ in the
$F\rightarrow\infty$ limit (see Appendix~\ref{sec:dynamical
Casimir effect in the weak mechanical driving regime}
and~\ref{sec:Semi-classical treatment for the strong mechanical
driving}). Hence, for a weak-$F$ drive, the very small
$\average{a_{s}^{\dag}a_{s}}_{\rm ss}$ leads to
$g_{s}^{\left(2\right)}\!\left(0\right)\gg1$. This corresponds to
strong photon bunching. In the special case of $F=0$, the $a_{s}$
mode cannot be excited although the two-photon driving still
exists, and as a consequence, the
$g_{s}^{\left(2\right)}\left(0\right)$ correlation cannot be
observed. We also find that with increasing the driving $F$, the
$g_{s}^{\left(2\right)}\!\left(0\right)$ correlation decreases and
then, as suggested above, approaches its lower bound equal to $1$.
These features confirm that the photons are bunched, as required.

So far, we have assumed a model with a squeezed-vacuum bath. To
avoid using such a bath and simplify the model, we now consider
the limit of $\Omega\ll\Delta$. In this limit, the effective
Hamiltonian is $\widetilde{H}_{\rm eff}$, as given above. In the
absence of the squeezed-vacuum bath, the $a$ mode is coupled to a
vacuum bath, and the master equation is the same as given in
    Eq.~(\ref{eq:master-equation}), but with $a_{s}\mapsto
    a$. We find that the noise induced by squeezing the cavity, which includes thermal noise
    $\propto\sinh^{2}\left(r\right)$ and two-photon correlation
    noise $\propto\sinh\left(2r\right)$, becomes strongly suppressed,
    even when there is no squeezed-vacuum bath. The DCE dynamics of the simplified model is therefore similar to what we have already demonstrated for the model that includes a squeezed-vacuum bath. Such a similarity can be made closer by decreasing the ratio
    $\Omega/\Delta$, but at the expense of the DCE radiation strength. In the limit of $\Omega\ll\Delta$, the background noise is $\approx0$, so that all the photons
    radiated from the cavity can be thought of as the DCE photons. For realistic parameters
$g_{0}=10\gamma_{m}$, $F=15\gamma_{m}$ and $\Omega/\Delta=0.1$, we
could obtain $\average{a^{\dag}a}_{\rm ss}\approx1.8\times10^{-3}$
at resonance ($\omega_{m}=\omega_{d}=2\omega_{s}$). This results
in an output flux $\approx2.0\times10^{4}$ photons per second for
$\kappa/2\pi=2$~MHz. This radiation can be measured using
single-photon detectors.

\section{Possible implementations}

As an example, we consider an {\it LC} superconducting circuit
with a micromechanical membrane (see
Appendix~\ref{sec:Possible-implementation-with-superconducting-circuits}
for details). In this device, the {\it LC} circuit is used to form
a single-mode microwave cavity. The mechanical motion of the
membrane modulates the capacitance of the {\it LC} circuit, and
thus the cavity frequency. In order to squeeze the cavity mode, an
additional tunable capacitor is embedded into the device. Its
cosine-wave modulation serves as a two-photon driving for the
cavity mode. The squeezed-vacuum reservoir can be generated
through an {\it LC} circuit with a tunable capacitor, or through a
Josephson parametric
amplifier~\cite{murch2013reduction,toyli2016resonance}.

Alternatively, our proposal can be implemented in an optical
system such as a whispering-gallery-mode (WGM) microresonator
coupled to a mechanical breathing
mode~\cite{kippenberg2005analysis,schliesser2006radiation,fiore2011storing,dong2012optomechanical,verhagen2012quantum,shen2016experimental,monifi2016optomechanically}.
The WGM microresonator made from nonlinear crystals exhibits
strong optical
nonlinearities~\cite{furst2011quantum,sedlmeir2017polarization,trainor2018selective},
which is the essential requirement for squeezing. The
squeezed-vacuum reservoir for the optical cavity can be prepared
by pumping a nonlinear medium, e.g., periodically-poled ${\rm
KTiOPO}_{4}$ (PPKTP) crystal, in a
cavity~\cite{ast2013high,serikawa2016creation,vahlbruch2016detection,schnabel2017squeezed}.

\section{Conclusions}

We have introduced a method for how to observe the mechanical DCE
in an optomechanical system. The method eliminates the problematic
need for an extremely high mechanical-oscillation frequency and an
ultrastrong single-photon optomechanical coupling. Thus, it paves
an experimentally feasible path to observing quantum radiation
from a moving mirror. Our method can be interpreted in the
laboratory frame as mechanically-induced two-photon hyper-Raman
scattering, an anti-Stokes process of scattering a driving photon
pair into a higher energy mode by absorbing a phonon. For the
absorbed phonon, its annihilation indicates the creation of a real
photon pair out of the quantum vacuum in the squeezed frame. We
have also showed a surprising result: that the squeezing, which
additionally induces and amplifies the DCE, can be extremely weak.
Note that in this case, the unconventional DCE can be considered
somehow similar to unconventional photon blockade
(UPB)~\cite{flayac2017unconventional}. Indeed, UPB is induced by a
nonlinearity, which can be extremely small. Finally, we expect
that the approach presented here could find diverse applications
in theoretical and experimental studies of quantum vacuum
radiation.

\begin{acknowledgments}
S.S. acknowledges the Army Research Office (ARO) (Grant No.
W911NF1910065). F.N. is supported in part by the: MURI Center for
Dynamic Magneto-Optics via the Air Force Office of Scientific
Research (AFOSR) (FA9550-14-1-0040), Army Research Office (ARO)
(Grant No. Grant No. W911NF-18-1-0358), Asian Office of Aerospace
Research and Development (AOARD) (Grant No. FA2386-18-1-4045),
Japan Science and Technology Agency (JST) (via the Q-LEAP program,
and the CREST Grant No. JPMJCR1676), Japan Society for the
Promotion of Science (JSPS) (JSPS-RFBR Grant No. 17-52-50023, and
JSPS-FWO Grant No. VS.059.18N), the RIKEN-AIST Challenge Research
Fund, the Foundational Questions Institute (FQXi), and the NTT PHI
Labs.
\end{acknowledgments}

\section*{APPENDICES}

\appendix 

\setcounter{equation}{0} \setcounter{figure}{0}
\setcounter{table}{0} \makeatletter
\renewcommand{\thefigure}{A\arabic{figure}}

\section{Optomechanical master equation, effective Hamiltonian, and off-resonant signal-to-noise ratio}
\label{sec:Optomechanical master equation}

\subsection{Optomechanical master equation}
In order to evaluate the steady-state behavior of the system, its
interaction with the environment needs to be described carefully.
In our proposal for observing the DCE, we parametrically squeeze
the cavity mode. Related methods have been used to enhance the
light-matter interaction in optomechanical
systems~\cite{lu2015squeezed,lemonde2016enhanced} and in cavity
electrodynamics systems~\cite{qin2018exponentially,
    leroux2018enhancing}. This can make the squeezed-cavity-mode
(SCM) frequency comparable to the mechanical frequency, so that
the mechanically induced DCE can be observed in a common
optomechanical setup without the need for an ultra-high mechanical
frequency and an ultrastrong single-photon optomechanical
coupling. However, the squeezing can also introduce undesired
noise, including thermal noise and two-photon correlation, into
the cavity. We can remove them by coupling a squeezed-vacuum bath
to the bare-cavity mode. In this section, we give a detailed
derivation of the master equation when the bare-cavity mode is
coupled to a squeezed-vacuum bath and the mechanical mode is
coupled to a thermal bath. We show that the noise induced by
squeezing the cavity can be completely eliminated.

To begin with, we consider the Hamiltonian for the interaction
between the system and the baths, which is given by
\begin{equation}
H_{\rm bath}=H_{\rm bath}^{0}+H_{\rm bath}^{c}+H_{\rm bath}^{m},
\end{equation}
where
\begin{align}
H_{\rm
bath}^{0}&=\sum_{l}\nu_{l}\left[t_{c}^{\dag}\left(\nu_{l}\right)t_{c}\left(\nu_{l}\right)
+t_{m}^{\dag}\left(\nu_{l}\right)t_{m}\left(\nu_{l}\right)\right],\\
H_{\rm
bath}^{c}&=\sum_{l}\lambda_{c}\left(\nu_{l}\right)\left[a^{\dag}t_{c}\left(\nu_{l}\right)
+t_{c}^{\dag}\left(\nu_{l}\right)a\right],\\
H_{\rm
bath}^{m}&=\sum_{l}\lambda_{m}\left(\nu_{l}\right)\left[b^{\dag}t_{m}\left(\nu_{l}\right)
+t_{m}^{\dag}\left(\nu_{l}\right)b\right].
\end{align}
Here, $H_{\rm bath}^{0}$ is the free Hamiltonian of the baths,
with $t_{c/m}\left(\nu_{l}\right)$ the annihilation operators for
the cavity and mechanical bath modes of frequency $\nu_{l}$, and
$H_{\rm bath}^{c/m}$ represent the couplings of the cavity and the
mechanical resonator to their baths, with  the coupling strengths
$\lambda_{c/m}\left(\nu_{l}\right)$ depending on the frequency
$\nu_{l}$. To derive the master equation, we first switch into the
frame rotating at
\begin{equation}
H_{0}=\omega_{L}a^{\dag}a/2+H_{\rm bath}^{0},
\end{equation}
to introduce the SCM using the Bogoliubov transformation
$a_{s}=\cosh\left(r\right)a+\sinh\left(r\right)a^{\dag}$. Then, we
again switch into the frame rotating at $H_{\rm
    CD}=\omega_{s}a_{s}^{\dag}a_{s}$, with
$\omega_{s}=\sqrt{\Delta^2-\Omega^2}$ being the SCM frequency,
where $\Delta=\omega_{c}-\omega_{L}/2$ is the detuning between the
bare-cavity frequency $\omega_{c}$ and the half-frequency,
$\omega_{L}/2$, of  the two-photon driving, and $\Omega$ is the
two-photon driving amplitude. The couplings between the system and
the baths are, accordingly, transformed to
\begin{align}
H_{\rm bath}^{c}\left(t\right)&=a\left(t\right)T_{c}^{\dag}\left(t\right)+a^{\dag}\left(t\right)T_{c}\left(t\right),\\
H_{\rm
bath}^{m}\left(t\right)&=b\left(t\right)T_{m}^{\dag}\left(t\right)+b^{\dag}\left(t\right)T_{m}\left(t\right).
\end{align}
Here, we have defined
\begin{align}
\label{seq:time-dependent-a}
a\left(t\right)&=\exp\left(-i\omega_{L}t/2\right)\exp\left(iH_{\rm CD}t\right)a\exp\left(-iH_{\rm CD}t\right),\\
b\left(t\right)&=\exp\left(-i\omega_{m}t\right)b,\\
T_{c}\left(t\right)&=\sum_{\nu_{l}}\lambda_{c}\left(\nu_{l}\right)t_{c}\left(\nu_{l}\right)\exp\left(-i\nu_{l}t\right),\\
T_{m}\left(t\right)&=\sum_{\nu_{l}}\lambda_{m}\left(\nu_{l}\right)t_{m}\left(\nu_{l}\right)\exp\left(-i\nu_{l}t\right).
\end{align}

Following the standard procedure in Ref.~\cite{scully1997book}
and, then, returning to the frame rotating at $H_{0}$, we can
obtain the following master equation expressed, in terms of the
$a_{s}$ mode, as
\begin{align}\label{seq:full-master-equation}
\frac{d}{dt}\rho\left(t\right)=\;&i\left[\rho\left(t\right),H\right]\nonumber\\
&-\frac{\kappa}{2}\left(N+1\right)\mathcal{L}\left(a_{s}\right)\rho\left(t\right)
-\frac{\kappa}{2}N\mathcal{L}\left(a_{s}^{\dag}\right)\rho\left(t\right)\nonumber\\
&+\frac{\kappa}{2}M\mathcal{L}^{\prime}\left(a_{s}\right)\rho\left(t\right)
+\frac{\kappa}{2}M^{*}\mathcal{L}^{\prime}\left(a_{s}^{\dag}\right)\rho\left(t\right)\nonumber\\
&-\frac{\gamma_{m}}{2}\left(n_{\rm
th}+1\right)\mathcal{L}\left(b\right)\rho\left(t\right)
-\frac{\gamma_{m}}{2}n_{\rm
th}\mathcal{L}\left(b^{\dag}\right)\rho\left(t\right),
\end{align}
where the Lindblad superoperators are defined by
\begin{align}
\mathcal{L}\left(o\right)\rho\left(t\right)
&=o^{\dag}o\rho\left(t\right)-2o\rho\left(t\right)o^{\dag}+\rho\left(t\right)o^{\dag}o,\\
\mathcal{L}^{\prime}\left(o\right)\rho\left(t\right)
&=oo\rho\left(t\right)-2o\rho\left(t\right)o+\rho\left(t\right)oo,
\end{align}
and  $N$, $M$ are given, respectively, by
\begin{align}
\label{seq:thermal-nosie-N}
N=&\cosh^{2}\left(r\right)\sinh^{2}\left(r_{e}\right)+\sinh^{2}\left(r\right)\cosh^{2}\left(r_{e}\right)\nonumber\\
&+\frac{1}{2}\sinh\left(2r\right)\sinh\left(2r_{e}\right)\cos\left(\theta_{e}\right),\\
\label{seq:two-photon-correlation-M}
M=&\left[\sinh\left(r\right)\cosh\left(r_{e}\right)
+\exp\left(-i\theta_{e}\right)\cosh\left(r\right)\sinh\left(r_{e}\right)\right]\nonumber\\
&\times\left[\cosh\left(r\right)\cosh\left(r_{e}\right)+\exp\left(i\theta_{e}\right)
\sinh\left(r\right)\sinh\left(r_{e}\right)\right],
\end{align}
corresponding to the thermal noise and two-photon correlation, and
where
\begin{align}
\kappa&=2\pi d_{c}\left(\omega_{L}/2\right)\lambda_{c}^{2}\left(\omega_{L}/2\right),\\
\gamma_{m}&=2\pi
d_{m}\left(\omega_{m}\right)\lambda_{m}^{2}\left(\omega_{m}\right),
\end{align}
represent, respectively, the cavity and mechanical decay rates,
with $d_{c}\left(\omega_{L}/2\right)$ being the density of states
for the cavity bath at frequency $\omega_{L}/2$, and
$d_{m}\left(\omega_{m}\right)$ being the density of states for the
mechanical bath at frequency $\omega_{m}$. Moreover, $n_{\rm
th}=\left[\exp\left(\omega_{m}/k_{B}T\right)-1\right]^{-1}$ is the
equilibrium phonon occupation at temperature $T$.

Note that, to derive the master equation in
Eq.~(\ref{seq:full-master-equation}), we have assumed that the
central frequency of the squeezed-vacuum bath is equal to half the
two-photon driving frequency. In addition, we have made the
following approximations,
\begin{align}
d_{c}\left(\omega_{L}/2\pm\omega_{s}\right)&\approx
d_{c}\left(\omega_{L}/2\right),\\
\lambda_{c}\left(\omega_{L}/2\pm\omega_{s}\right)&\approx
\lambda_{c}\left(\omega_{L}/2\right).
\end{align}
This is because, in our case, the SCM frequency $\omega_{s}$ is
tuned to be comparable to the mechanical frequency $\omega_{m}$
($\sim$~MHz). Thus, it is much smaller than the two-photon driving
frequency $\omega_{L}$ (of the order of GHz for microwave light or
even THz for optical light).

According to Eqs.~(\ref{seq:thermal-nosie-N}) and
(\ref{seq:two-photon-correlation-M}), we can have $N=M=0$ for
$r_{e}=r$ and $\theta_{e}=\pm n\pi$ ($n=1,3,5,\cdots$), and thus,
we have,
\begin{align}\label{seq:master-equation-in-terms-of-the-squeezed-mode}
\frac{d}{dt}\rho\left(t\right)=\;&i\left[\rho\left(t\right),H\right]-\frac{\kappa}{2}\mathcal{L}\left(a_{s}\right)\rho\left(t\right)\nonumber\\
&-\frac{\gamma_{m}}{2}\left(n_{\rm
th}+1\right)\mathcal{L}\left(b\right)\rho\left(t\right)
-\frac{\gamma_{m}}{2}n_{\rm
th}\mathcal{L}\left(b^{\dag}\right)\rho\left(t\right).
\end{align}
We find from
Eq.~(\ref{seq:master-equation-in-terms-of-the-squeezed-mode}) that
the squeezing-induced noise is completely eliminated, so that the
$a_{s}$ mode is equivalently coupled to the thermal vacuum bath.
As we demonstrate below, eliminating this noise can ensure that
the background noise is zero for the SCM excitation spectrum in
the squeezed frame, and as a result, the background noise of the
output-photon flux spectrum in the original laboratory frame only
originates from photons contained in the squeezed vacuum. This
minimizes the background noise for the observation of the DCE, and
thus enables the DCE to be observed more clearly in experiments.

\begin{figure}[h]
    \centering
    \includegraphics[width=8.5cm]{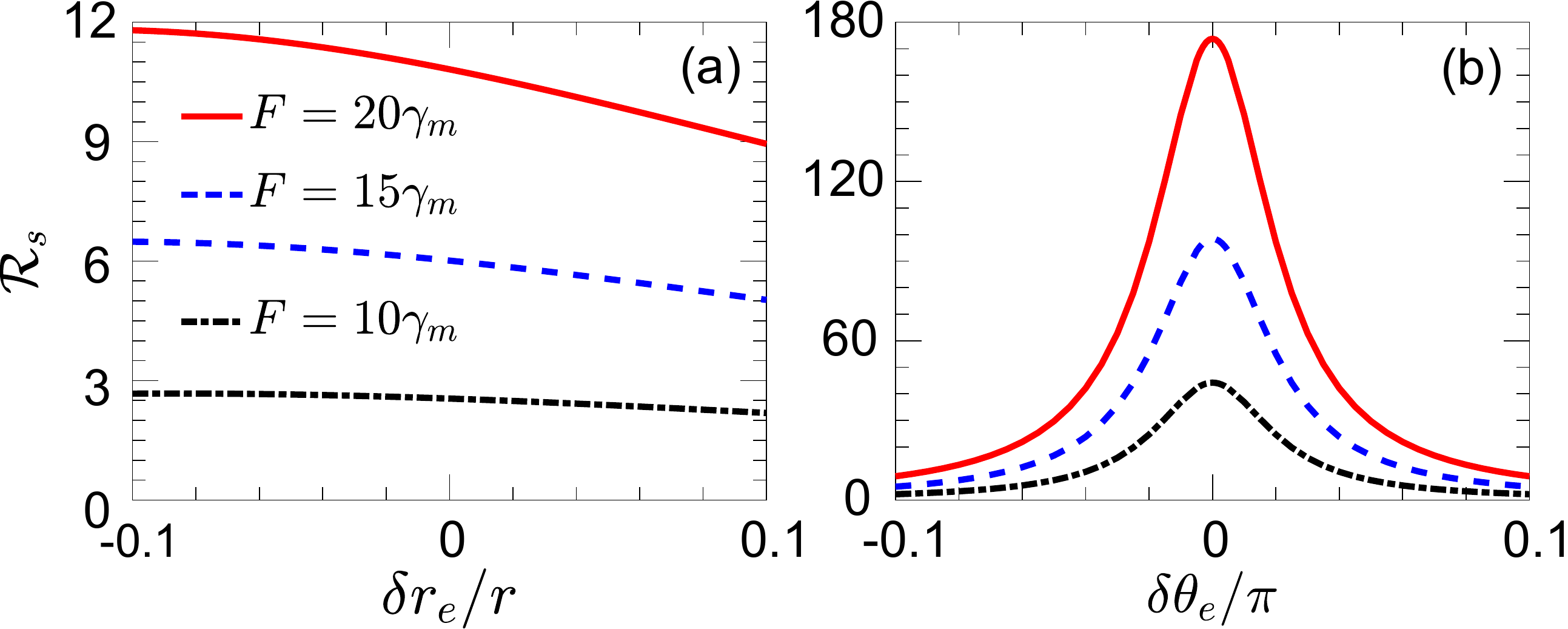}
    \caption{Signal-to-noise ratio $\mathcal{R}_{s}$ versus (a) $\delta r_{e}=r_{e}-r$ and (b) $\delta \theta_{e}=\theta_{e}-\pi$ for the driving strength $F=10\gamma_{m}$, $15\gamma_{m}$, and $20\gamma_{m}$. In (a), we set $\theta_{e}=1.1\pi$, and in (b) $r_{e}=1.1r$. In both plots, we set $g_{0}=10\gamma_{m}$, $n_{\rm th}=0$, $\omega_{m}=\omega_{d}=2\omega_{s}=10^{4}\gamma_{m}$, $\Delta=\omega_{m}$, and $\kappa=500\gamma_{m}$.}\label{sfig_SNR_squeezed_frame}
\end{figure}

When the conditions $r=r_{e}$ and $\theta_{e}=\pm n\pi$
($n=1,3,5,\cdots$) are not perfectly satisfied, the
squeezing-induced noise cannot be eliminated completely (i.e.,
$N\neq0$ and $M\neq0$). However, according to the master equation
in Eq. (\ref{seq:full-master-equation}), such imperfections do not
affect the occurrence of the DCE. They only cause some noises. To
quantify this undesired effect, we use the signal-to-noise ratio
defined as
\begin{equation}
\mathcal{R}_{s}=\frac{\langle a_{s}^{\dag}a_{s}\rangle_{\rm
ss}^{F\neq0}-\langle a_{s}^{\dag}a_{s}\rangle_{\rm
ss}^{F=0}}{\langle a_{s}^{\dag}a_{s}\rangle_{\rm ss}^{F=0}},
\end{equation}
where $\langle a_{s}^{\dag}a_{s}\rangle_{\rm ss}^{F=0}$ ($\langle
a_{s}^{\dag}a_{s}\rangle_{\rm ss}^{F\neq0}$) is the steady state
$\langle a_{s}^{\dag}a_{s}\rangle$ when $F=0$  ($F\neq0$), and the
subscript ``ss" stands for ``steady state". We plot
$\mathcal{R}_{s}$ in Fig.~\ref{sfig_SNR_squeezed_frame}, according
to the master equation given in
Eq.~(\ref{seq:full-master-equation}) but replacing $H\mapsto
H_{\rm eff}$. In this figure, we assume that $r_{e}=r+\delta
r_{e}$ and $\theta_{e}=\pi+\delta\theta_{e}$. In the perfect case
of $N=M=0$,  $\mathcal{R}_{s}\rightarrow\infty$ because $\langle
a_{s}^{\dag}a_{s}\rangle_{\rm ss}^{F=0}=0$. Thus, we find in
Fig.~\ref{sfig_SNR_squeezed_frame} that the noise induced by
imperfect parameters reduces the ratio $\mathcal{R}_{s}$. However,
we also find that with increasing the driving $F$, the noise
becomes smaller compared to the DCE signal, such that it can even
be neglected for sufficiently strong $F$.

\subsection{Effective Hamiltonian}
\begin{figure}[t]
    \centering
    \includegraphics[width=8.5cm]{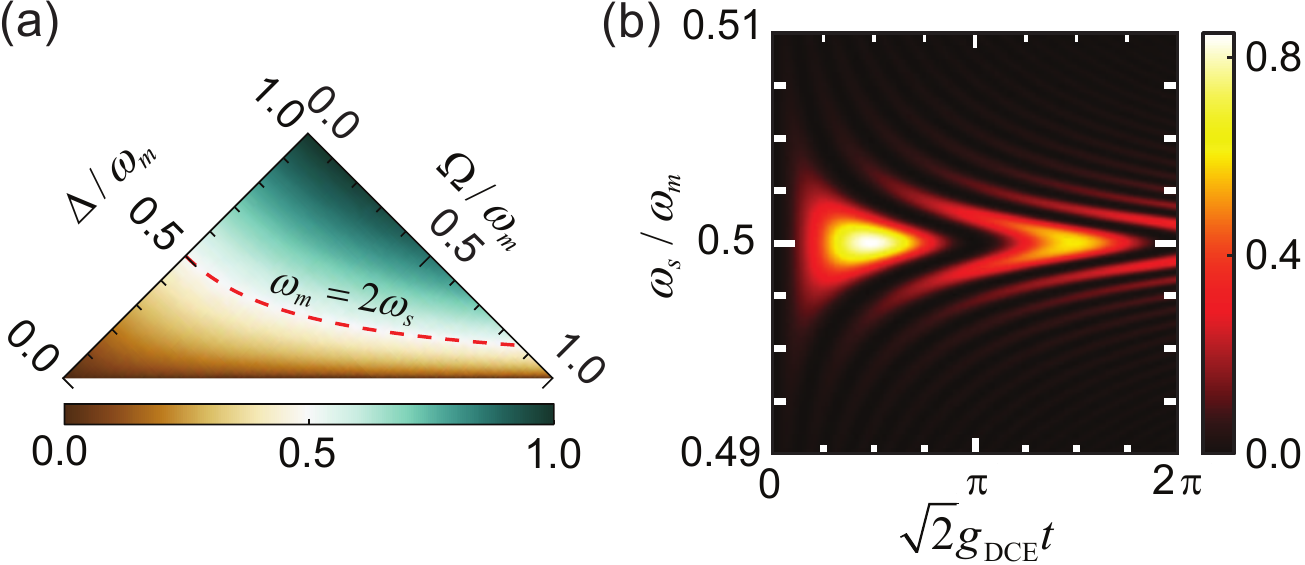}
    \caption{(a) Squeezed-cavity-mode (SCM) frequency
        $\omega_{s}$ as a function of the parametric-driving detuning
        $\Delta$ and strength $\Omega$. The dashed curve represents the
        $\omega_{m}=2\omega_{s}$ case. Here, in order for the system to be
        stable, we need to have $\Delta>\Omega$. (b) Parametric energy conversion between the mechanical and squeezed cavity modes, obtained from a numerical solution of the master equation in Eq.~(\ref{seq:master-equation-in-terms-of-the-squeezed-mode}). Under the time evolution, a single phonon can simultaneously excite two SCM photons. Here, the initial state is $|0_{\rm s},1\rangle$ and the desired state is $|2_{\rm s},0\rangle$, where the first number in the ket refers to the SCM photon number and the second to the mechanical phonon number. Moreover, we set $g_{0}=80\gamma_{m}$, $n_{\rm th}=0$, $\omega_{m}=10^{3}g_{0}$, $\Delta=\omega_{m}$, $\kappa=10\gamma_{m}$, and $F=0$.}\label{sfig:onephonontotwophoton}
\end{figure}

\begin{figure}[b]
    \centering
    \includegraphics[width=7.0cm]{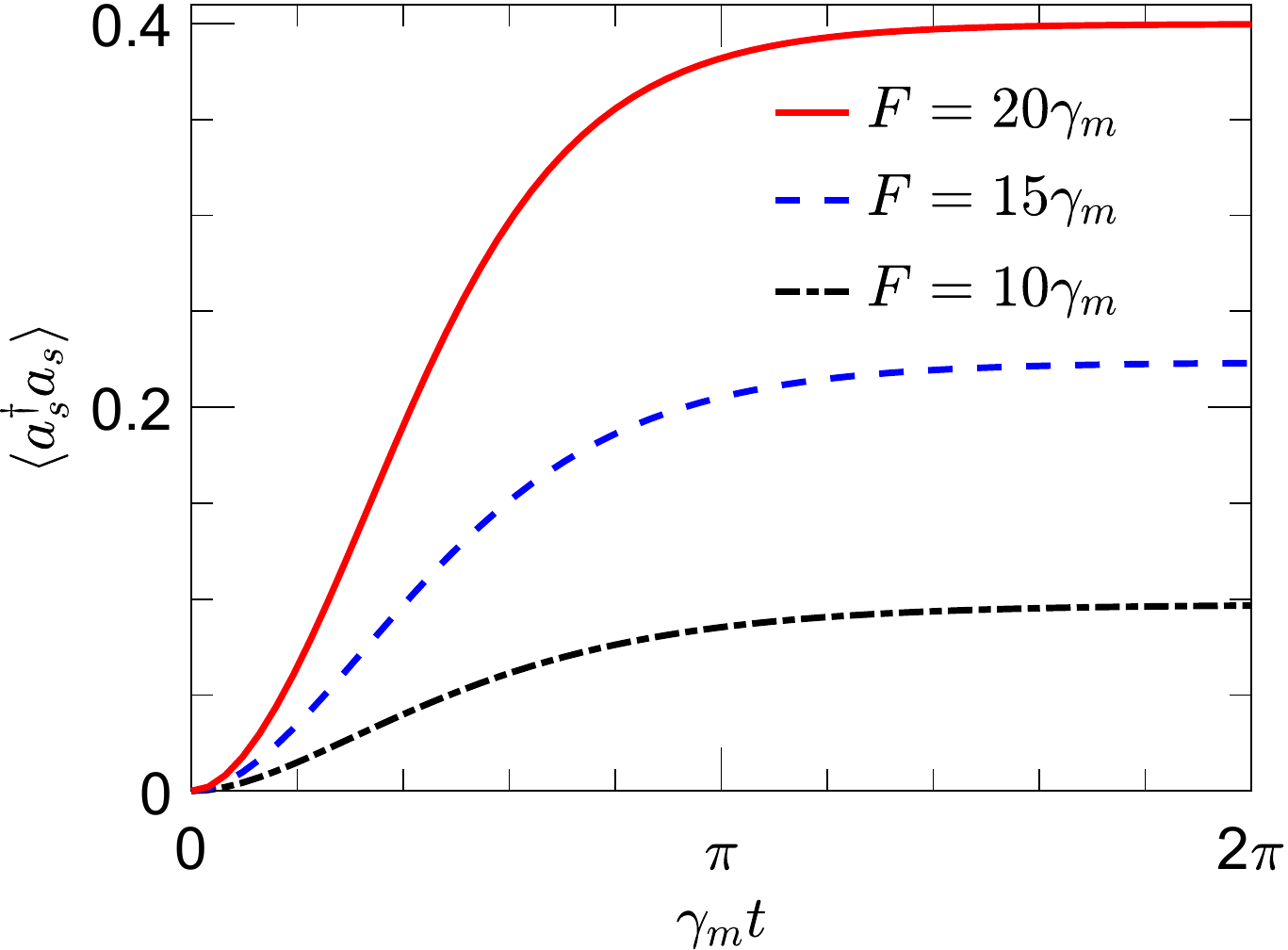}
    \caption{Time evolution of $\langle a_{s}^{\dag}a_{s}\rangle$ for the driving strength $F=10\gamma_{m}$, $15\gamma_{m}$, and $20\gamma_{m}$, calculated from the master equation in Eq.~(\ref{seq:effective-master-equation-in-terms-of-the-squeezed-mode}). Here, the initial state is assumed to be $|0_{\rm s},0\rangle$. Furthermore, we assume that $g_{0}=10\gamma_{m}$, $n_{\rm th}=0$, $\omega_{m}=\omega_{d}=2\omega_{s}=10^{4}\gamma_{m}$, $\Delta=\omega_{m}$, and $\kappa=500\gamma_{m}$.}\label{sfig_dynamics_evolution}
\end{figure}

The Hamiltonian in Eqs.~(\ref{seq:full-master-equation}) and
(\ref{seq:master-equation-in-terms-of-the-squeezed-mode}) is
expressed, in terms of the $a_{s}$ mode, as
\begin{align}\label{seq:full-Hamiltonian-in-squeezed-frame}
H=\;&\omega_{s}a_{s}^{\dag}a_{s}+\omega_{m}b^{\dag}b-g_{\rm OM}a_{s}^{\dag}a_{s}\left(b+b^{\dag}\right)\nonumber\\
&+g_{\rm DCE}\left(a_{s}^{2}+a_{s}^{\dag2}\right)\left(b+b^{\dag}\right)\nonumber\\
&+\frac{F}{2}\left[\exp\left(i\omega_{d}t\right)b+\exp\left(-i\omega_{d}t\right)b^{\dag}\right],
\end{align}
where $g_{\rm OM}=g_{0}\cosh\left(2r\right)$ and $g_{\rm
    DCE}=g_{0}\sinh\left(2r\right)/2$, with
$r=\left(1/4\right)\ln\left[\left(\Delta+\Omega\right)/\left(\Delta-\Omega\right)\right]$
being the squeezing parameter of the cavity. In
Fig.~\ref{sfig:onephonontotwophoton}(a) we plot $\omega_{s}$ as a
function of $\Delta$ and $\Omega$, and find that the resonance
condition $\omega_{m}=2\omega_{s}$, for a parametric coupling
between SCM and mechanical mode, can be achieved with
experimentally modest parameters. The Hamiltonian $H$ essentially
describes the optomechanical system where the boundary condition
of a squeezed field is modulated by the mechanical motion of a
driven mirror. In the limit $\left\{\omega_{s},
\omega_{m},\omega_{d}\right\}\gg\left\{g_{\rm OM}, g_{\rm DCE},
F\right\}$, we can apply the rotating-wave approximation, such
that the coherent dynamics of the system is governed by the
following effective Hamiltonian,
\begin{align}\label{seq:JC-like-Hamiltonian-in-terms-of-as}
H_{\rm eff}=\;&\Delta_{s}a_{s}^{\dag}a_{s}+\Delta_{m}b^{\dag}b\nonumber\\
&+g_{\rm
DCE}\left(a_{s}^{2}b^{\dag}+a_{s}^{\dag2}b\right)+\frac{F}{2}\left(b+b^{\dag}\right),
\end{align}
where $\Delta_{s}=\omega_{s}-\omega_{d}/2$ and
$\Delta_{m}=\omega_{m}-\omega_{d}$. The master equation in
Eq.~(\ref{seq:master-equation-in-terms-of-the-squeezed-mode}) is
then reduced to
\begin{align}\label{seq:effective-master-equation-in-terms-of-the-squeezed-mode}
\frac{d}{dt}\rho\left(t\right)=\;&i\left[\rho\left(t\right),H_{\rm eff}\right]-\frac{\kappa}{2}\mathcal{L}\left(a_{s}\right)\rho\left(t\right)\nonumber\\
&-\frac{\gamma_{m}}{2}\left(n_{\rm
th}+1\right)\mathcal{L}\left(b\right)\rho\left(t\right)
-\frac{\gamma_{m}}{2}n_{\rm
th}\mathcal{L}\left(b^{\dag}\right)\rho\left(t\right).
\end{align}
We find, according to
Eq.~(\ref{seq:JC-like-Hamiltonian-in-terms-of-as}), that the
coupling of the states $|0_{\rm s},1\rangle$ and $|2_{\rm
s},0\rangle$, where the first number in the ket refers to the SCM
photon number and the second one to the mechanical phonon number,
is given by
\begin{equation}
g_{|0_{\rm s},1\rangle\leftrightarrow|2_{\rm
s},0\rangle}=\sqrt{2}g_{\rm DCE}.
\end{equation}
In the squeezed frame, this means that under the time evolution,
one phonon can be converted into two photons, and vice versa, at
resonance $\omega_{m}=2\omega_{s}$. To confirm such a state
conversion, we perform numerics, as shown in
Fig.~\ref{sfig:onephonontotwophoton}(b). Specifically, we use the
master equation in
Eq.~(\ref{seq:master-equation-in-terms-of-the-squeezed-mode}) to
calculate the fidelity, $\mathcal{F}=\langle2_{\rm s},0|\rho_{\rm
actual}\left(t\right)|2_{\rm s},0\rangle$, where $\rho_{\rm
actual}\left(t\right)$ is the actual state. It is seen in
Fig.~\ref{sfig:onephonontotwophoton}(b) that we have the expected
state conversion between light and mechanics, and there is a
maximum conversion at resonance. Note that owing to the presence
of the cavity and mechanical losses, the maximum conversion
fidelity decreases with time.

To describe the dynamics of the DCE further, we plotted the time
evolution of $\langle a_{s}^{\dag}a_{s}\rangle$ in the presence of
the driving $F$ in Fig.~\ref{sfig_dynamics_evolution}.  We find
that $\langle a_{s}^{\dag}a_{s}\rangle$ increases with time and
then gradually approaches its stationary value. For an
experimental parameter $\gamma_{m}\approx200$~Hz in
Ref.~\cite{teufel2011sideband}, the stationary state is reached
within a time $\approx5/\gamma_{m}\approx25$~ms.

\begin{figure}[h]
    \centering
    \includegraphics[width=7.0cm]{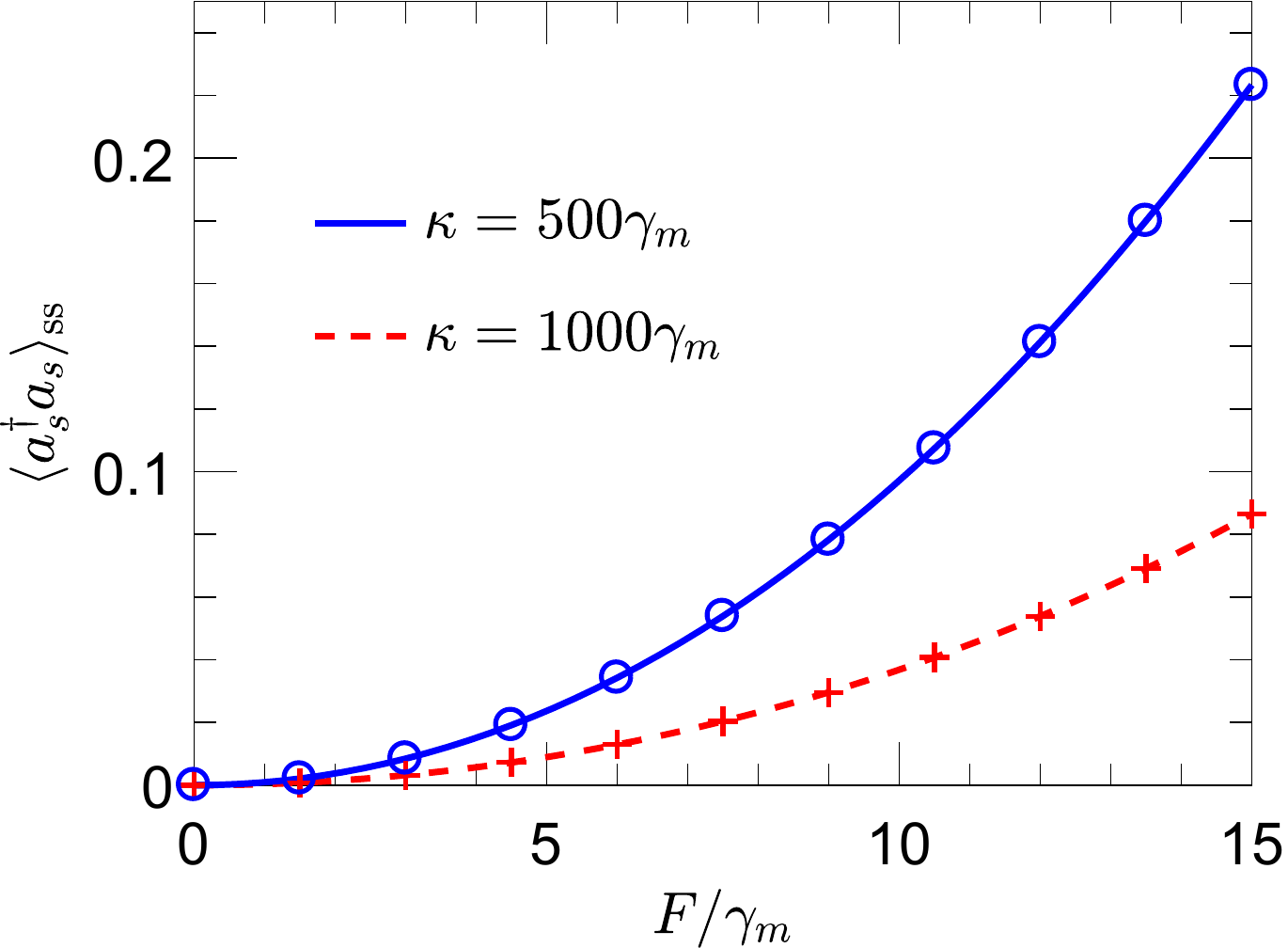}
    \caption{Effects of the high-frequency component $H_{\rm high}$ on the photon number $\langle a_{s}^{\dag}a_{s}\rangle_{\rm ss}$. The master equation used for curves is given in Eq.~(\ref{seq:effective-master-equation-in-terms-of-the-squeezed-mode}), and for symbols is given in Eq.~(\ref{seq:master-equation-in-terms-of-the-squeezed-mode}) but with $H$ in Eq.~(\ref{seq:averaged_full_hamiltonian}). Here, we set $g_{0}=10\gamma_{m}$, $n_{\rm th}=0$, $\omega_{m}=\omega_{d}=2\omega_{s}=10^{4}\gamma_{m}$, and $\Delta=\omega_{m}$.}\label{sfig_no_RWA_steady_state}
\end{figure}

In Eq.~(\ref{seq:JC-like-Hamiltonian-in-terms-of-as}), we made the
rotating-wave approximation and neglected the high-frequency
component
\begin{align}
H_{\rm high}=&-g_{\rm OM}a_{s}^{\dag}a_{s}\left[\exp\left(-i\omega_{d}t\right)b+\exp\left(i\omega_{d}t\right)b^{\dag}\right]\nonumber\\
&+g_{\rm
DCE}\left[\exp\left(-i2\omega_{d}t\right)a_{s}^{2}b+\exp\left(i2\omega_{d}t\right)a_{s}^{\dag2}b^{\dag}\right].
\end{align}
In typical situations, $\left\{g_{\rm OM}, g_{\rm
DCE}\right\}\ll\omega_{d}$, which allows a time-averaging
treatment of $H_{\rm high}$ using the formalism of
Ref.~\cite{gamel2010time}. After a straightforward calculation,
the behavior of $H_{\rm high}$ can be approximated, at resonance
$\omega_{m}=\omega_{d}=2\omega_{s}$ (i.e.,
$\Delta_{s}=\Delta_{m}=0$), as
\begin{align}\label{seq:time-average-Hamiltonian}
H_{\rm high}\approx H_{\rm TA}=&-\frac{g_{\rm OM}^{2}}{\omega_{m}}\left(a_{s}^{\dag}a_{s}\right)^{2}-\frac{g_{\rm DCE}^{2}}{2\omega_{m}}\big[a_{s}^{\dag2}a_{s}^{2}\nonumber\\
&+2\left(2a_{s}^{\dag}a_{s}+1\right)b^{\dag}b+2\left(2a_{s}^{\dag}a_{s}+1\right)\big].
\end{align}
The Hamiltonian $H$ is, accordingly, transformed to
\begin{equation}\label{seq:averaged_full_hamiltonian}
H\approx H_{\rm eff}+H_{\rm TA}.
\end{equation}
For realistic parameters, the couplings $g_{\rm OM}$ and $g_{\rm
DCE}$ are three orders of magnitude lower than $\omega_{m}$. We
can find from Eq.~(\ref{seq:time-average-Hamiltonian}) that the
high-frequency term $H_{\rm high}$ can be neglected, compared to
the low-frequency term $H_{\rm eff}$. To confirm this, in
Fig.~\ref{sfig_no_RWA_steady_state} we numerically calculated
$\langle a_{s}^{\dag}a_{s}\rangle_{\rm ss}$ using the
low-frequency term $H_{\rm eff}$ and the full Hamiltonian $H$
given in Eq.~(\ref{seq:averaged_full_hamiltonian}), respectively.
By comparing these, we find an excellent agreement, and the
high-frequency term $H_{\rm high}$ can be safely neglected, as
expected.

\subsection{Off-resonant signal-to-noise ratio}

\begin{figure}[t]
    \centering
    \includegraphics[width=8.5cm]{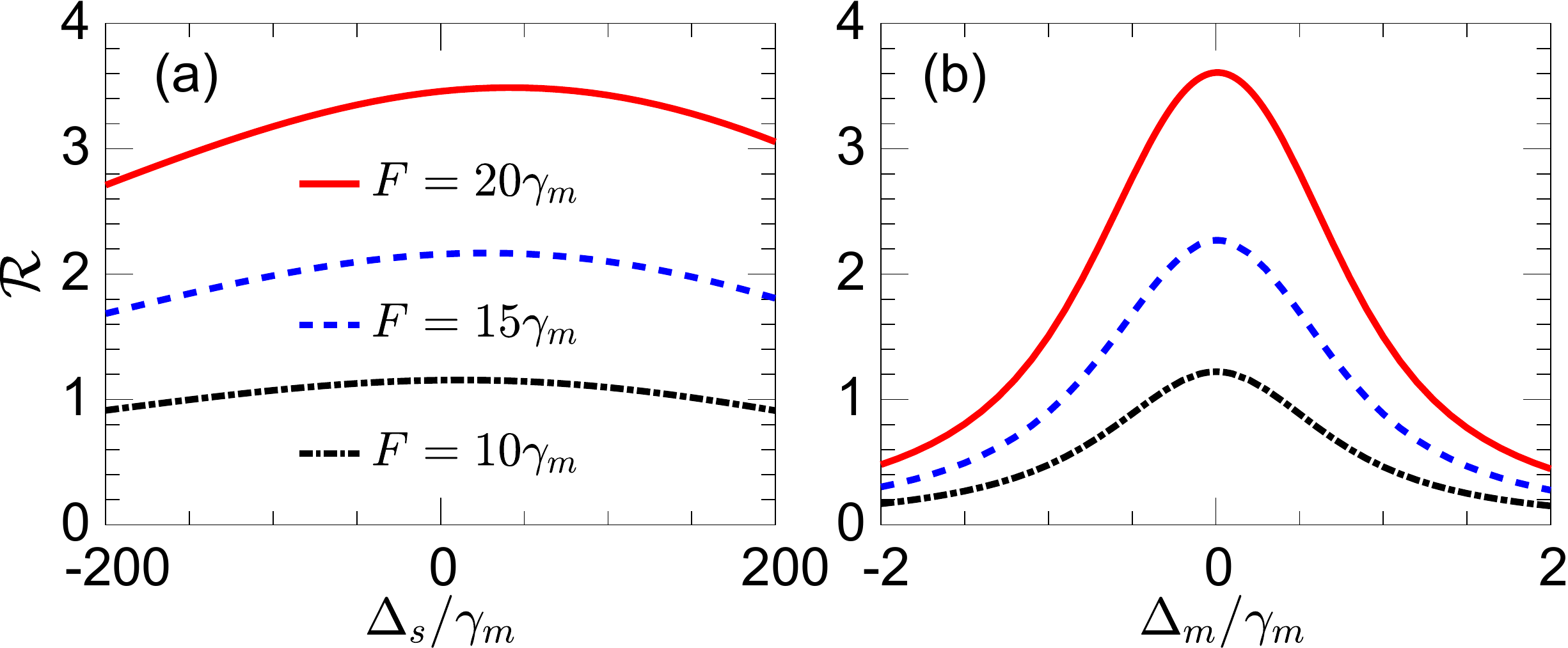}
    \caption{Signal-to-noise ratio  $\mathcal{R}$ versus detunings (a) $\Delta_{s}$ and (b) $\Delta_{m}$ for the driving strength $F=10\gamma_{m}$, $15\gamma_{m}$, and $20\gamma_{m}$. The master equation used here is given in Eq.~(\ref{seq:effective-master-equation-in-terms-of-the-squeezed-mode}). In (a), we set $\Delta_{m}=0.2\gamma_{m}$ and in (b) $\Delta_{s}=10\gamma_{m}$. In both plots, we set $g_{0}=10\gamma_{m}$, $n_{\rm th}=0$, $\kappa=500\gamma_{m}$, and $\sinh^{2}\left(r\right)=0.5$.}\label{sfig_SNR_detuning}
\end{figure}

In the main article, the signal-to-noise ratio $\mathcal{R}$ is
discussed at resonance $\omega_{m}=\omega_{d}=2\omega_{s}$ (i.e.,
$\Delta_{s}=\Delta_{m}=0$). We now discuss the ratio $\mathcal{R}$
in the off-resonance case where $\Delta_{s}\neq0$ and
$\Delta_{m}\neq0$. We plot the ratio $\mathcal{R}$ as a function
of the detunings $\Delta_{s}$ and $\Delta_{m}$ in
Fig.~\ref{sfig_SNR_detuning}. There, the results are obtained by
numerically integrating the master equation in
Eq.~(\ref{seq:effective-master-equation-in-terms-of-the-squeezed-mode}).
We find that the ratio $\mathcal{R}$ decreases with the detuning
$\Delta_{s}$ or $\Delta_{m}$, but increases with the force $F$.
Note that the DCE photons are the scattered photon pairs via
two-photon hyper-Raman scattering. As a result, their frequency
$\omega_{s}+\omega_{L}/2$ is different from the noise-photon
frequency $\omega_{L}/2$. This means that if standard techniques
of Raman spectroscopy are used, the noise can then be filtered
out. Therefore, the signal can still be resolved even if $R < 1$.

\section{Dynamical Casimir effect in the mechanical weak-driving regime}
\label{sec:dynamical Casimir effect in the weak mechanical driving
regime}

In our main article, we have studied the steady-state behavior
associated with the DCE, by numerically integrating the master
equation in
Eq.~(\ref{seq:effective-master-equation-in-terms-of-the-squeezed-mode})~\cite{johansson2012qutip,
johansson2013qutip2}. To study the DCE further, an analytical
understanding for the mechanical weak driving is given in this
section. Here, we only focus on the resonance situation where
$\omega_{m}=\omega_{d}=2\omega_{s}$.

\begin{figure*}[t]
    \centering
    \includegraphics[width=17.0cm]{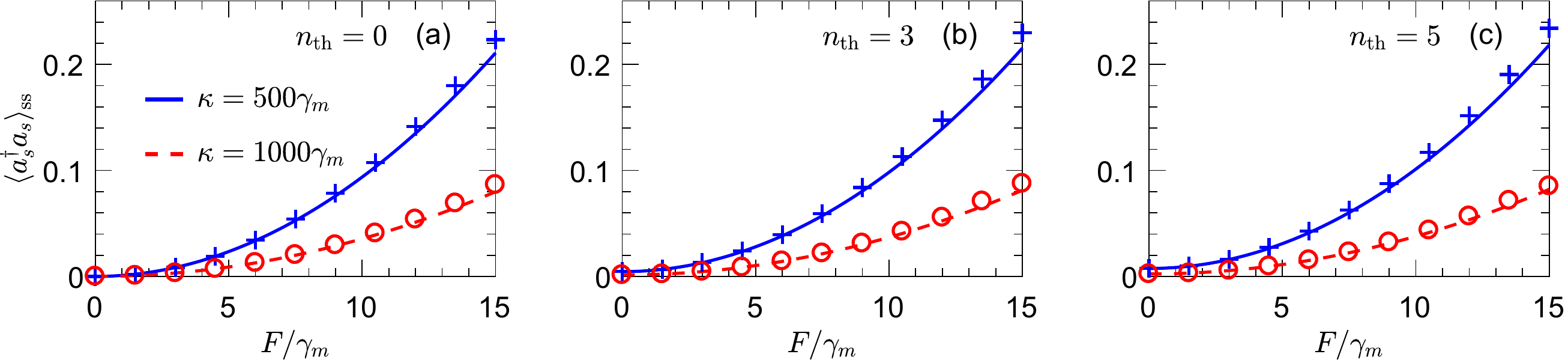}
    \caption{Photon number $\langle a_{s}^{\dag}a_{s}\rangle_{\rm ss}$ versus mechanical driving $F$ for (a) $n_{\rm th}=0$, (b) $n_{\rm th}=3$, and (c) $n_{\rm th}=5$. Curves are analytical results, while symbols are numerical simulations of the master equation in Eq.~(\ref{seq:effective-master-equation-in-terms-of-the-squeezed-mode}). Here, we set $g_{0}=10\gamma_{m}$, $\omega_{m}=\omega_{d}=2\omega_{s}=10^{4}\gamma_{m}$, and $\Delta=\omega_{m}$.}\label{sfig-intracavity-squeezed-photon-number}
\end{figure*}

\begin{figure*}[tbph]
    \centering
    \includegraphics[width=17.0cm]{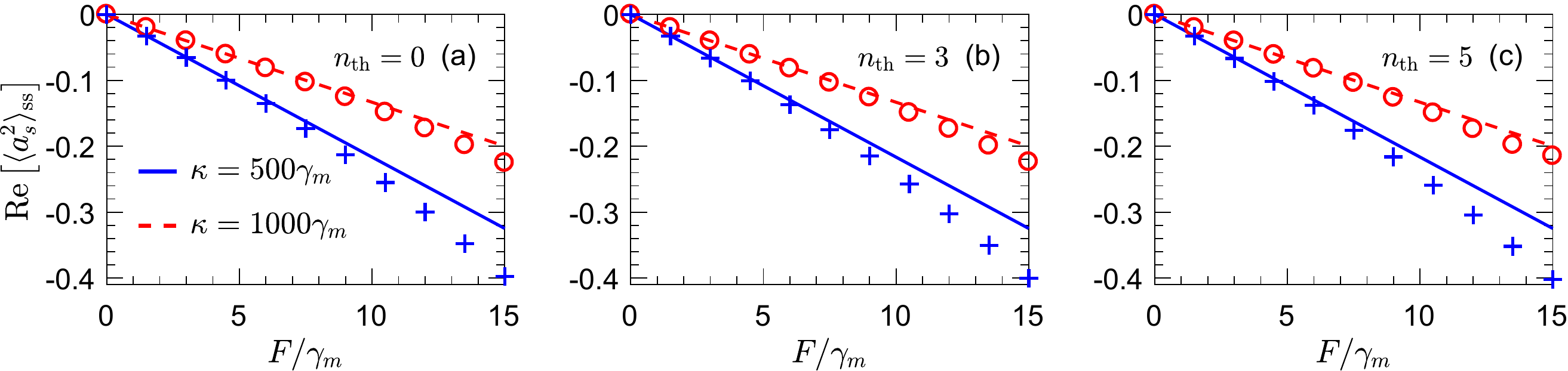}
    \caption{Real part of the correlation function
        $\average{a_{s}^{2}}_{\rm ss}$ versus mechanical driving $F$ for
        (a) $n_{\rm th}=0$, (b) $n_{\rm th}=3$, and (c) $n_{\rm th}=5$.
        We assumed the same curve correspondences and the same parameters
        as in
        Fig.~\ref{sfig-intracavity-squeezed-photon-number}.}\label{sfig-intracavity-asas}
\end{figure*}
Let us now derive the steady-state SCM photon number $\langle
a_{s}^{\dag}a_{s}\rangle_{\rm ss}$. To begin, we consider the
master equation in
Eq.~(\ref{seq:effective-master-equation-in-terms-of-the-squeezed-mode}).
The involved equations of motion are given, respectively, by
\begin{align}
\label{seq:differential-equation01}
\frac{d}{dt}\langle a_{s}^{\dag}a_{s}\rangle=-&4g_{\rm DCE}\;{\rm Im}\left[\average{a_{s}^{2}b^{\dag}}\right]-\kappa\langle a_{s}^{\dag}a_{s}\rangle,\\
\label{seq:differential-equation02}
\frac{d}{dt}\average{a_{s}^{2}}=&-i2g_{\rm DCE}\left(2\average{a_{s}^{\dag}a_{s}b}+\average{b}\right)-\kappa\average{a_{s}^{2}},\\
\label{seq:differential-equation03}
\frac{d}{dt}\average{b}=&-i\left(g_{\rm DCE}\average{a_{s}^{2}}+\frac{F}{2}\right)-\frac{\gamma_{m}}{2}\average{b},\\
\label{seq:differential-equation04}
\frac{d}{dt}\average{b^{\dag}b}=\;&2g_{\rm DCE}\;{\rm Im}\left[\average{a_{s}^{2}b^{\dag}}\right]-F{\rm Im}\left[\average{b}\right]\nonumber\\
&-\gamma_{m}\average{b^{\dag}b}
+\gamma_{m}n_{\rm th},\\
\label{seq:differential-equation05}
\frac{d}{dt}\average{a_{s}^{2}b^{\dag}}=\;&i\bigg(g_{\rm DCE}\average{a_{s}^{\dag 2}a_{s}^{2}}-4g_{\rm DCE}\average{a_{s}^{\dag}a_{s}b^{\dag}b}\nonumber\\
&+\frac{F}{2}\average{a_{s}^{2}}-2g_{\rm DCE}\average{b^{\dag}b}\bigg)\nonumber\\
&-\left(\kappa+\frac{\gamma_{m}}{2}\right)\average{a_{s}^{2}b^{\dag}},\\
\label{seq:differential-equation06}
\frac{d}{dt}\average{a_{s}^{\dag}a_{s}b}=\;&ig_{\rm DCE}\left(2\average{a_{s}^{2}b^{\dag}b}-\average{a_{s}^{\dag}a_{s}^{3}}-2\average{a_{s}^{\dag 2}b^{2}}\right)\nonumber\\
&-i\frac{F}{2}\average{a_{s}^{\dag}a_{s}}-\left(\kappa+\frac{\gamma_{m}}{2}\right)\average{a_{s}^{\dag}a_{s}b}.
\end{align}
Here, ${\rm Im}\left[z\right]$ represents the imaginary part of
$z$. In fact, owing to the parametric coupling, the Hamiltonian in
Eq.~(\ref{seq:JC-like-Hamiltonian-in-terms-of-as}) leads to an
infinite set of differential equations, which may not be
analytically solved. Thus, in order to obtain an analytical
result, we neglect the higher-order correlation terms, that is:
$\average{a_{s}^{\dag 2}a_{s}^{2}}$,
$\average{a_{s}^{\dag}a_{s}b^{\dag}b}$,
$\average{a_{s}^{2}b^{\dag}b}$, $\average{a_{s}^{\dag}a_{s}^{3}}$,
and $\average{a_{s}^{\dag 2}b^{2}}$. This approximation is valid
for a weak driving $F$, as shown below. In such an approximation,
the coupled differential equations
~(\ref{seq:differential-equation01})--(\ref{seq:differential-equation06})
construct a closed set, so in the steady state we have
\begin{align}
\label{seq:coupledequation1}
0\approx&-4g_{\rm DCE}\;{\rm Im}\left[\average{a_{s}^{2}b^{\dag}}_{\rm ss}\right]-\kappa\langle a_{s}^{\dag}a_{s}\rangle_{\rm ss},\\
0\approx&-i2g_{\rm DCE}\left(2\average{a_{s}^{\dag}a_{s}}_{\rm ss}+\average{b}_{\rm ss}\right)-\kappa\average{a_{s}^{2}}_{\rm ss},\\
0\approx&-i\left(g_{\rm DCE}\average{a_{s}^{2}}_{\rm ss}+\frac{F}{2}\right)-\frac{\gamma_{m}}{2}\average{b}_{\rm ss},\\
0\approx\;&2g_{\rm DCE}\;{\rm Im}\left[\average{a_{s}^{2}b^{\dag}}_{\rm ss}\right]-F\;{\rm Im}\left[\average{b}_{\rm ss}\right]-\gamma_{m}\average{b^{\dag}b}_{\rm ss}\nonumber\\
&+\gamma_{m}n\left(\omega_{m},T\right),\\
0\approx\;&i\left(\frac{F}{2}\average{a_{s}^{2}}-2g_{\rm DCE}\average{b^{\dag}b}_{\rm ss}\right)-\left(\kappa+\frac{\gamma_{m}}{2}\right)\average{a_{s}^{2}b^{\dag}}_{\rm ss},\\
\label{seq:coupledequation6}
0\approx&-i\frac{F}{2}\average{a_{s}^{\dag}a_{s}}_{\rm
ss}-\left(\kappa+\frac{\gamma_{m}}{2}\right)\average{a_{s}^{\dag}a_{s}b}_{\rm
ss}.
\end{align}
By solving this closed set of equations, the steady-state SCM
photon number is found to be
\begin{equation}\label{seq:steady-state-intracavity-photon-number}
\average{a_{s}^{\dag}a_{s}}_{\rm ss}\approx\frac{4\gamma_{m}g_{\rm
DCE}^{2}}{\kappa\left(2g_{\rm
DCE}^{2}+\gamma_{0}^{2}\right)}\left[\frac{\kappa
F^{2}}{2\gamma_{m}\left(2g_{\rm DCE}+\gamma_{0}^{2}\right)}+n_{\rm
th}\right],
\end{equation}
where $\gamma_{0}=\sqrt{\kappa\gamma_{m}/2}$.
Equation~(\ref{seq:steady-state-intracavity-photon-number}) shows
that $\average{a_{s}^{\dag}a_{s}}_{\rm ss}$ includes two physical
contributions: one from the mechanical driving and the other from
the thermal noise. Furthermore, we also find a quadratic increase
in $\average{a_{s}^{\dag}a_{s}}_{\rm ss}$ with the driving $F$. To
confirm this analytical expression, in
Fig.~\ref{sfig-intracavity-squeezed-photon-number} we compare it
with exact numerical simulations of the master equation in
Eq.~(\ref{seq:effective-master-equation-in-terms-of-the-squeezed-mode}).
It is seen that the analytical predictions are in good agreement
with the exact numerical results, especially for weak $F$.

According to the Bogoliubov transformation, the steady-state
intracavity-photon number $\average{a^{\dag}a}_{\rm ss}$ in the
original laboratory frame is given in
Eq.~(\ref{eq:cavity_photon_number}). Then, the steady-state
output-photon flux is given in Eq.~(\ref{eq:output_flux}).
\begin{figure*}[tbph]
    \centering
    \includegraphics[width=17.0cm]{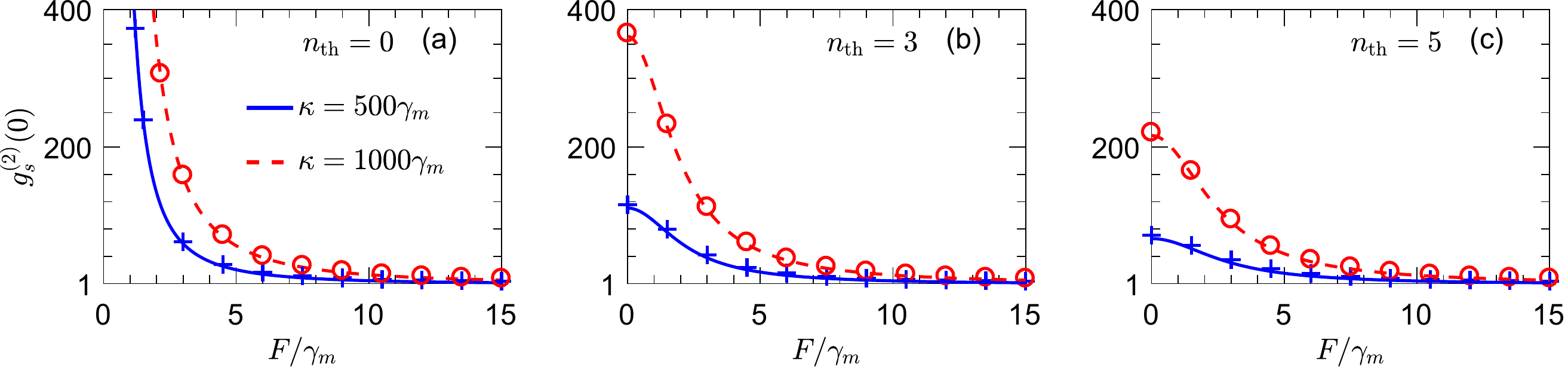}
    \caption{Correlation function
        $g^{\left(2\right)}_{s}\!\left(0\right)$ versus mechanical driving
        $F$ for (a) $n_{\rm th}=0$, (b) $n_{\rm th}=3$, and (c) $n_{\rm
            th}=5$. We assumed the same curve correspondences and the same
        parameters as in
        Fig.~\ref{sfig-intracavity-squeezed-photon-number}.}\label{sfig-g2}
\end{figure*}
To obtain $\Phi_{\rm out}$ analytically, the physical quantities
$\average{a_{s}^{\dag}a_{s}}_{\rm ss}$ and ${\rm
    Re}\left[\average{a_{s}^{2}}_{\rm ss}\right]$ are involved, as
shown in Eq.~(\ref{eq:DCE_signal}). The steady-state SCM photon
number, $\average{a_{s}^{\dag}a_{s}}_{\rm ss}$, is given in
Eq.~(\ref{seq:steady-state-intracavity-photon-number}), and
further is numerically confirmed in
Fig.~\ref{sfig-intracavity-squeezed-photon-number}. From the
closed set of the steady-state equations given in
Eqs.~(\ref{seq:coupledequation1})--(\ref{seq:coupledequation6}),
we can straightforwardly find
\begin{equation}\label{seq:DCE-induced-asas}
\average{a_{s}^{2}}_{\rm ss}=-\frac{g_{\rm DCE}}{2g_{\rm
DCE}^{2}+\gamma_{0}^{2}}F.
\end{equation}
It shows that $|{\rm Re}\left[\average{a_{s}^{2}}_{\rm
ss}\right]|$ increases linearly with $F$ but is independent of the
thermal mechanical noise. This behavior is also numerically
confirmed in Fig.~\ref{sfig-intracavity-asas}, showing a good
agreement especially for the weak driving $F$. Note that the
derivation of the analytical results and their numerical
confirmations originates from neglecting the higher-order
correlation terms. In order to exactly describe
$\average{a_{s}^{2}}$, such higher-order correlations should be
included. By combining
Eqs.~(\ref{seq:steady-state-intracavity-photon-number})--(\ref{seq:DCE-induced-asas}),
the steady-state output-photon flux can be analytically expressed
as
\begin{align}\label{seq:steady-state-output}
\Phi_{\rm out}=&\frac{4\gamma_{m}g_{\rm DCE}^{2}}{\kappa\left(2g_{\rm DCE}^{2}+\gamma_{0}^{2}\right)}\left[\frac{\kappa }{2\gamma_{m}\left(2g_{\rm DCE}+\gamma_{0}^{2}\right)}F^{2}+n_{\rm th}\right]\nonumber\\
&\times\cosh\left(2r\right)\nonumber\\
&+\frac{g_{\rm DCE}}{2g_{\rm
DCE}^{2}+\gamma_{0}^{2}}\sinh\left(2r\right)F+\kappa\sinh^{2}\left(r\right).
\end{align}
We find from Eq.~(\ref{seq:steady-state-output}) that, by
increasing the mechanical driving $F$, the DCE-induced photon flux
$\Phi_{\rm DCE}$ becomes stronger quadratically, but at the same
time, the background-noise photon flux $\Phi_{\rm BGN}$ remains
unchanged. Therefore, the increase in the total photon flux
$\Phi_{\rm out}$ with $F$ can be considered as a signature of the
mechanical-motion induced DCE.

In the DCE process, the photons are emitted in pairs, and
therefore, they could exhibit photon
bunching~\cite{johansson2010dynamical,stassi2013spontaneous,macri2018nonperturbative}.
The essential parameter quantifying this property is the
equal-time second-order correlation function, defined in
Eq.~(\ref{eq:g2_correlation}). We now derive this second-order
correlation function. The equation of motion for
$\average{a_{s}^{\dag 2}a_{s}^{2}}$ is given by
\begin{align}
\frac{d}{dt}\average{a_{s}^{\dag 2}a_{s}^{2}}=&-4g_{\rm DCE}\left\{2{\rm Im}\left[\average{a_{s}^{\dag}a_{s}^{3}b^{\dag}}\right]+{\rm Im}\left[\average{a_{s}^{2}b^{\dag}}\right]\right\}\nonumber\\
&-2\kappa\average{a_{s}^{\dag 2}a_{s}^{2}}.
\end{align}
We can neglect the term ${\rm
Im}\left[\average{a_{s}^{\dag}a_{s}^{3}b^{\dag}}\right]$ for the
weak driving $F$. Then, combining Eq.~(\ref{seq:coupledequation1})
yields
\begin{equation}\label{seq:second-order-correlation-weak}
g_{s}^{\left(2\right)}\!\left(0\right)\approx\frac{1}{2\average{a_{s}^{\dag}a_{s}}_{\rm
ss}}.
\end{equation}
In Fig.~\ref{sfig-g2}, we plot the
$g_{s}^{\left(2\right)}\!\left(0\right)$ correlation as a function
of the driving $F$. In this figure, we compare the analytical and
numerical results, and show an exact agreement. Owing to a very
small of $\average{a_{s}^{\dag}a_{s}}_{\rm ss}$ for the mechanical
weak  driving, $g_{s}^{\left(2\right)}\!\left(0\right)$ is very
large as shown in Fig.~\ref{sfig-g2}, which corresponds to large
photon bunching. With increasing the driving $F$, we also find
that the $g_{s}^{\left(2\right)}\!\left(0\right)$ correlation
decreases, and as demonstrated more explicitly in the
Appendix~\ref{sec:Semi-classical treatment for the strong
mechanical driving}, it would approach a lower bound equal to $1$,
thereby implying that the DCE radiation field becomes a coherent
state in the limit of the mechanical strong driving,
$F\rightarrow\infty$.

\section{Semi-classical treatment for the dynamical Casimir effect}
\label{sec:Semi-classical treatment for the strong mechanical
driving}

In Appendix~\ref{sec:dynamical Casimir effect in the weak
mechanical driving regime} we have analytically discussed the DCE
process when the mechanical driving $F$ is weak. There, the
higher-order correlations that arise from the parametric coupling
are neglected, and the resulting expressions can predict the
system behavior well. For strong-$F$ driving, all high-order
correlations should be included to exactly describe the system;
but in this case, finding solutions analytically or even
numerically becomes much more difficult. In order to investigate
the DCE in the strong-$F$ regime, in this section we employ a
semi-classical treatment~\cite{butera2019mechanical}. For
simplicity, but without loss of generality, here we assume that
the mechanical resonator is coupled to a zero-temperature bath.
For finite temperatures, the discussion below is still valid, as
long as the total number of phonons is much larger than the number
of thermal phonons.

\subsection{Excitation spectrum and output-photon flux spectrum in the steady state}

We again begin with the master equation in
Eq.~(\ref{seq:effective-master-equation-in-terms-of-the-squeezed-mode})
and, accordingly, obtain
\begin{align}
\label{seq:semiclassical-equation01}
\frac{d}{dt}\langle a_{s}^{\dag}a_{s}\rangle=&-4g_{\rm DCE}\;{\rm Im}\left[\average{a_{s}^{2}}\average{b}^{*}\right]-\kappa\langle a_{s}^{\dag}a_{s}\rangle,\\
\label{seq:semiclassical-equation02}
\frac{d}{dt}\average{a_{s}^{2}}=&-i2\Delta_{s}\average{a_{s}^{2}}\nonumber\\
&-i2g_{\rm DCE}\left(2\average{a_{s}^{\dag}a_{s}}+1\right)\average{b}-\kappa\average{a_{s}^{2}},\\
\label{seq:semiclassical-equation03}
\frac{d}{dt}\average{b}=&-i\left(\Delta_{m}\average{b}+g_{\rm
DCE}\average{a_{s}^{2}}+\frac{F}{2}\right)-\frac{\gamma_{m}}{2}\average{b}.
\end{align}
Here, we have made the semiclassical approximation, such that
$\average{a_{s}^{2}b^{\dag}}\approx\average{a_{s}^{2}}\average{b}^{*}$
and
$\average{a_{s}^{\dag}a_{s}b}\approx\average{a_{s}^{\dag}a_{s}}\average{b}$.
Under this approximation, the fluctuation correlation between the
cavity and the mechanical resonator is neglected. It is found that
Eqs.~(\ref{seq:semiclassical-equation01})--(\ref{seq:semiclassical-equation03})
construct a closed set.

\begin{figure*}[t]
    \centering
    \includegraphics[width=17.0cm]{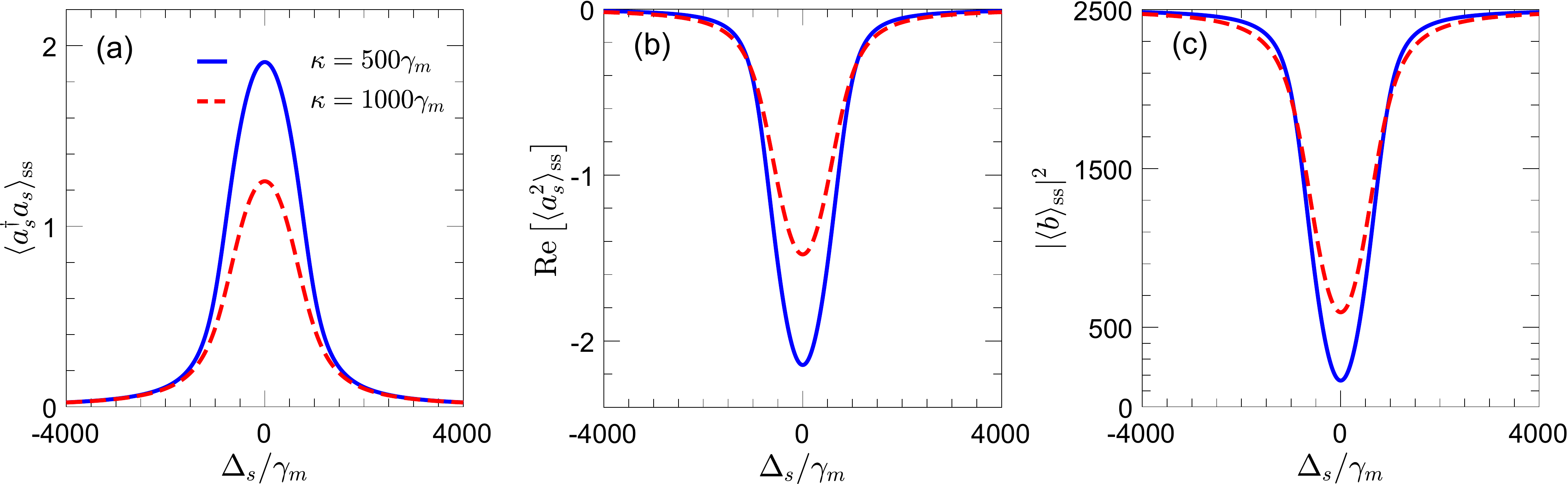}
\caption{(a) Photon number $\average{a_{s}^{\dag}a_{s}}_{\rm ss}$,
(b) real part of the correlation function
$\average{a_{s}^{2}}_{\rm ss}$, and (c) phonon number
$|\average{b}_{\rm ss}|^{2}$ versus detuning
$\Delta_{s}=\omega_{s}-\omega_{d}/2$ for $\kappa=500\gamma_{m}$
(solid curves) and $\kappa=1000\gamma_{m}$ (dashed curves). We
have assumed that $\omega_{m}=\omega_{d}$, $g_{0}=10\gamma_{m}$,
$F=50\gamma_{m}$, and
$\sinh^{2}\left(r\right)=0.5$.}\label{sfig:omega_m=omega_d}
\end{figure*}

\subsubsection{Excitation spectrum for resonant mechanical driving: $\omega_{m}=\omega_{d}$}

We first consider the case of a resonant mechanical driving (i.e.,
$\omega_{m}=\omega_{d}$). In this case, we have $\Delta_{m}=0$,
and the steady-state SCM photon number
$\average{a_{s}^{\dag}a_{s}}_{\rm ss}$ satisfies a cubic equation,
\begin{align}\label{seq:cubic-equation-01}
0=\;&g_{\rm DCE}^{4}\,x^{3}-g_{\rm DCE}^{2}\left(g_{\rm DCE}^{2}-\gamma_{0}^{2}\right)x^{2}\nonumber\\
&+\left[\frac{1}{4}\left(\Delta_{s}^{2}\gamma_{m}^{2}+\gamma_{0}^{4}\right)-g_{\rm DCE}^{2}\left(F^{2}+\gamma_{0}^{2}\right)\right]x\nonumber\\
&-\frac{1}{4}\left(\Delta_{s}^{2}\gamma_{m}^{2}+\gamma_{0}^{4}\right),
\end{align}
where $\gamma_{0}=\sqrt{\kappa\gamma_{m}/2}$ and
$x=2\average{a_{s}^{\dag}a_{s}}_{\rm ss}+1$. The solutions of such
a equation can be exactly obtained using the Cardano formula.
Then, the steady-state $\average{a_{s}^{2}}$ and $\average{b}$ are
given, respectively, by
\begin{align}
\label{seq:strong-steady-asas}
\average{a_{s}^{2}}_{\rm ss}&=-\frac{g_{\rm DCE}}{2g_{\rm DCE}^{2}\,x
+\gamma_{0}^{2}+i\Delta_{s}\gamma_{m}}Fx,\\
\label{seq:strong-steady-b} \average{b}_{\rm
ss}&=-\frac{\left(-2\Delta_{s}+i\kappa\right)}{2g_{\rm
DCE}^{2}\,x+\gamma_{0}^{2}+i\Delta_{s}\gamma_{m}}\frac{F}{2}.
\end{align}
For simplicity, we numerically solve the cubic
equation~(\ref{seq:cubic-equation-01}), and in
Fig.~\ref{sfig:omega_m=omega_d} we plot
$\average{a_{s}^{\dag}a_{s}}_{\rm ss}$, ${\rm
Re}\left[\average{a_{s}^{2}}_{\rm ss}\right]$, and
$|\average{b}_{\rm ss}|^{2}$ versus the detuning $\Delta_{s}$ for
$\kappa=500\gamma_{m}$ and $1000\gamma_{m}$. At large detunings,
the resonantly driven mechanical resonator is effectively
decoupled from the cavity mode. As a consequence, there is almost
no conversion of mechanical energy into photons. Thus at large
detunings, the mechanical phonon number $|\average{b}_{\rm
ss}|^{2}$ quickly approaches $\left(F/\gamma_m\right)^{2}$, i.e.,
the steady-state phonon number when the mechanical resonator is
completely uncoupled. Meanwhile, both the photon number
$\average{a_{s}^{\dag}a_{s}}_{\rm ss}$ and correlation function
$\average{a_{s}^{2}}_{\rm ss}$ are very close to zero. As the
detuning decreases, the effective parametric coupling between the
mechanical motion and the cavity mode increases, and the
parametric conversion from mechanical energy into photons is
accordingly enhanced. Such an energy conversion is maximized at
resonance $\omega_{m}=\omega_{d}=2\omega_{s}$. Thus, when
decreasing the detuning, both $\average{a_{s}^{\dag}a_{s}}_{\rm
ss}$ and $|{\rm Re}\left[\average{a_{s}^{2}}_{\rm ss}\right]|$
increase but $|\average{b}_{\rm ss}|^{2}$ decreases, as shown in
Fig.~\ref{sfig:omega_m=omega_d}. In particular,
$\average{a_{s}^{\dag}a_{s}}_{\rm ss}$ and $|{\rm
Re}\left[\average{a_{s}^{2}}_{\rm ss}\right]|$ reach their maximum
values at resonance, and at the same time, $|\average{b}_{\rm
ss}|^{2}$ reaches its minimum value. This behavior implies that
the photons are emitted by the mechanical resonator.

\begin{figure*}[tbph]
    \centering
    \includegraphics[width=17.0cm]{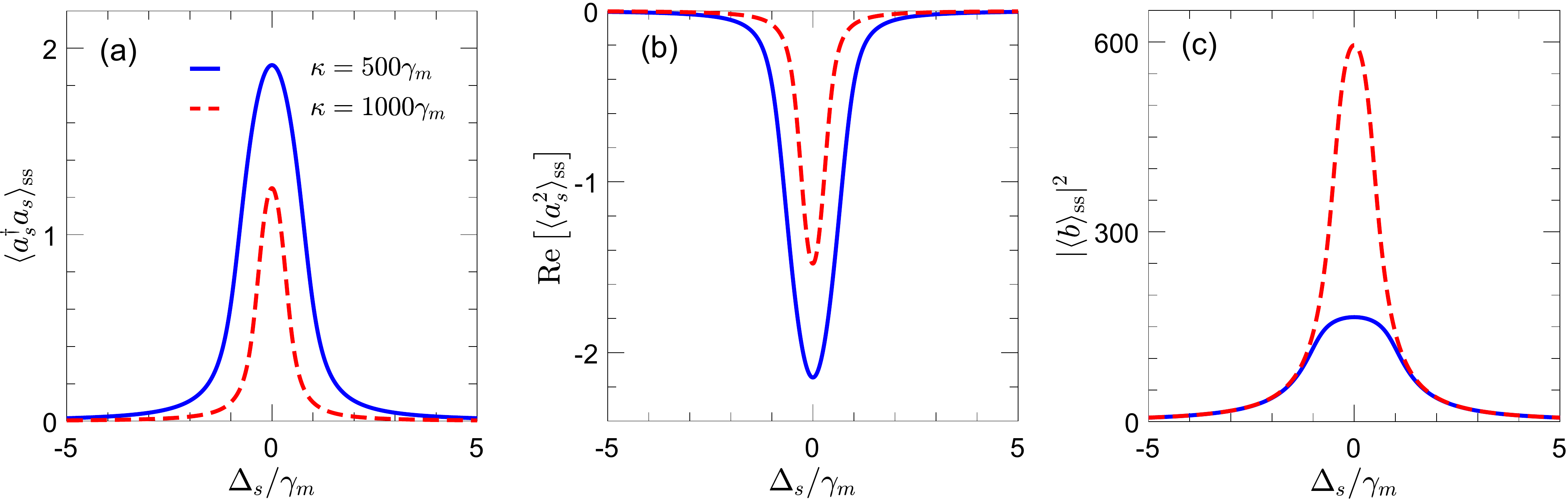}
    \caption{(a) Photon number $\average{a_{s}^{\dag}a_{s}}_{\rm ss}$, (b) real part of the correlation function $\average{a_{s}^{2}}_{\rm ss}$, and (c) phonon number $|\average{b}_{\rm ss}|^{2}$ (c) versus detuning $\Delta_{s}=\omega_{s}-\omega_{d}/2$ for $\kappa=500\gamma_{m}$ (solid curves) and $\kappa=1000\gamma_{m}$ (dashed curves). We have assumed that $\omega_{m}=2\omega_{s}$, $g_{0}=10\gamma_{m}$, $F=50\gamma_{m}$, and $\sinh^{2}\left(r\right)=0.5$.}\label{sfig:omega_m=2omega_s}
\end{figure*}

\subsubsection{Excitation spectrum for resonant parametric coupling: $\omega_{m}=2\omega_{s}$}

We next consider the case of a resonant parametric coupling (i.e.,
$\omega_{m}=2\omega_{s}$). In this case, we have
$\Delta_{m}=2\Delta_{s}=\Delta$, and the steady-state
$\average{a_{s}^{\dag}a_{s}}$ also satisfies a cubic equation
\begin{align}\label{seq:cubic-equation-02}
0=\;&g_{\rm DCE}^{4}\,x^{3}-g_{\rm DCE}^{2}\left(g_{\rm DCE}^{2}+\Delta^{2}-\gamma_{0}^{2}\right)x^{2}\nonumber\\
&+\bigg\{\frac{1}{4}\left[\left(\Delta^2-\gamma_{0}^{2}\right)^{2}
+\Delta^{2}\gamma_{1}^{2}\right]+g_{\rm DCE}^{2}\big(\Delta^2\nonumber\\
&-\gamma_{0}^{2}-F^{2}\big)\bigg\}x-\frac{1}{4}\left[\left(\Delta^{2}-\gamma_{0}^{2}\right)^{2}+\Delta^{2}\gamma_{1}^{2}\right],
\end{align}
where $\gamma_{1}=\kappa+\gamma_{m}/2$. This cubic equation can
also be exactly solved using the Cardano formula, and then the
steady-state $\average{a_{s}^{2}}$ and $\average{b}$ are given,
respectively, by
\begin{align}
\average{a_{s}^{2}}_{\rm ss}&=-\frac{g_{\rm DCE}}{2g_{\rm DCE}\,x-\Delta^{2}+\gamma_{0}^{2}+i\Delta\gamma_{1}}Fx,\\
\average{b}_{\rm ss}&=-\frac{\left(-\Delta+i\kappa\right)}{2g_{\rm
DCE}\,x-\Delta^{2}+\gamma_{0}^{2}+i\Delta\gamma_{1}}\frac{F}{2}.
\end{align}
We numerically solve the cubic
equation~(\ref{seq:cubic-equation-02}), and in
Fig.~\ref{sfig:omega_m=2omega_s}, we plot
$\average{a_{s}^{\dag}a_{s}}_{\rm ss}$, ${\rm
Re}\left[\average{a_{s}^{2}}_{\rm ss}\right]$, and
$|\average{b}_{\rm ss}|^{2}$ versus the detuning $\Delta_{s}$ for
$\kappa=500\gamma_{m}$ and $1000\gamma_{m}$. At large detunings,
the mechanical driving is effectively decoupled from the
mechanical resonator, so that almost no phonons are excited and
almost no photons are emitted. As the detuning decreases, the
mechanical phonon number increases, which strengthens the
parametric conversion from mechanical energy into photons, and in
turn, leads to an increase in the excited photon number. This
process is maximized at resonance
$\omega_{m}=\omega_{d}=2\omega_{s}$. Thus, we find, as shown in
Fig.~\ref{sfig:omega_m=2omega_s}, that with decreasing the
detuning, not only $\average{a_{s}^{\dag}a_{s}}_{\rm ss}$ and
$|{\rm Re}\left[\average{a_{s}^{2}}_{\rm ss}\right]|$ but also
$|\average{b}_{\rm ss}|^{2}$ increases, and that they
simultaneously reach their maximum values at resonance. This
behavior also implies that the photons are emitted by the
mechanical resonator.

\subsubsection{Output-photon flux spectrum for resonant mechanical driving and parametric coupling}

\begin{figure}[h]
    \centering
    \includegraphics[width=8.5cm]{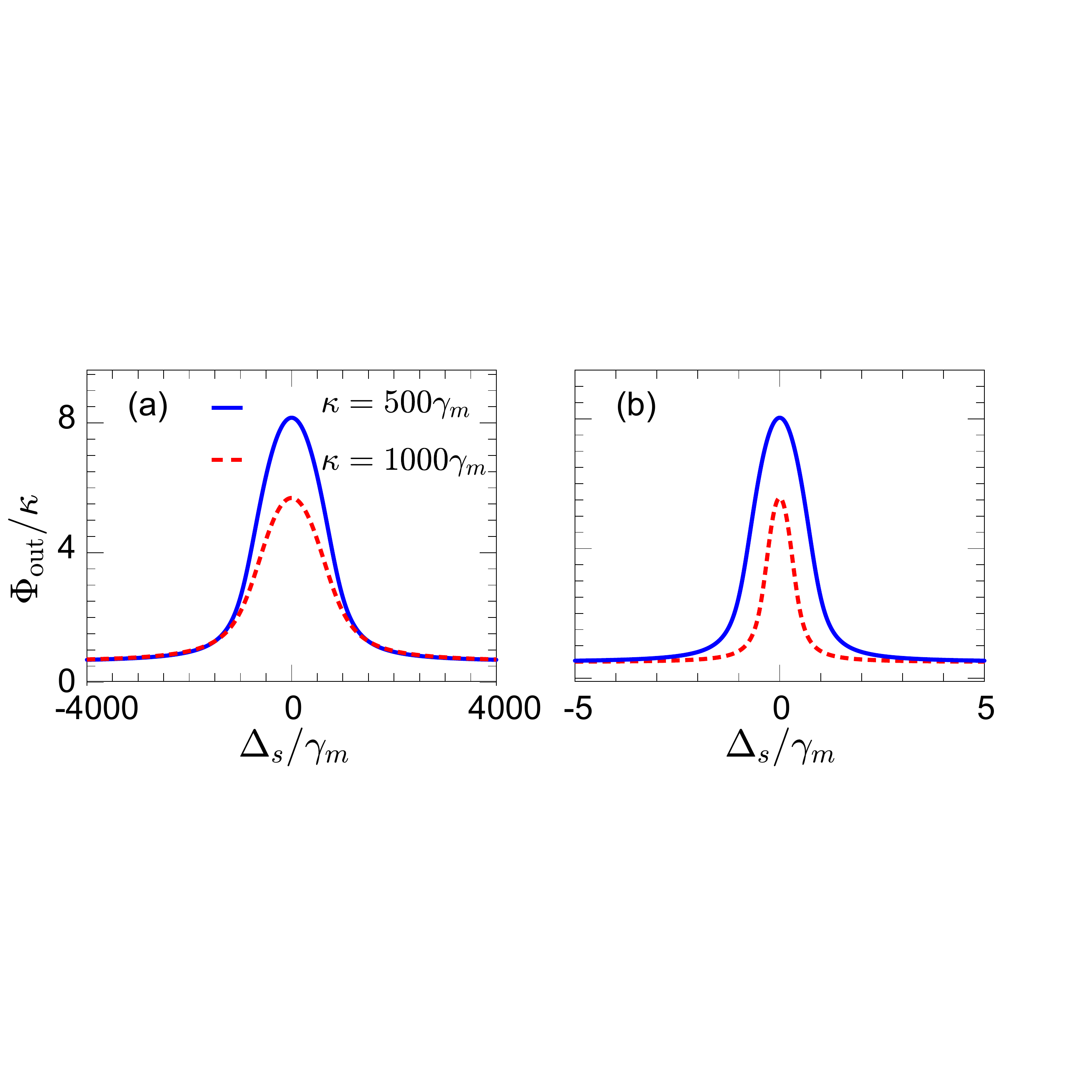}
    \caption{Steady-state output-photon flux $\Phi_{\rm out}$ as a function of the detuning $\Delta_{s}=\omega_{s}-\omega_{d}/2$ for $\kappa=500\gamma_{m}$ (solid curves) and $1000\gamma_{m}$ (dashed curves). We assumed that $\omega_{m}=\omega_{d}$ (resonant mechanical driving) in (a) and $\omega_{m}=2\omega_{s}$ (resonant parametric coupling) in (b). For both plots, we assumed that $g_{0}=10\gamma_{m}$, $F=50\gamma_{m}$, and $\sinh^{2}\left(r\right)=0.5$.}\label{sfig:output-photon-flux-strong}
\end{figure}

Having obtained $\average{a_{s}^{\dag}a_{s}}_{\rm ss}$ and
$\average{a_{s}^{2}}_{\rm ss}$ in the squeezed frame, we can,
according to the Bogoliubov transformation, calculate the
steady-state intracavity photon number $\average{a^{\dag}a}_{\rm
    ss}$ in the original laboratory frame, as given in
Eq.~(\ref{eq:cavity_photon_number}). Then we can calculate the
steady-state output-photon flux $\Phi_{\rm out}$ according to the
input-output relation, given in Eq.~(\ref{eq:output_flux}). We
plot the photon flux $\Phi_{\rm out}$ as a function of the
detuning $\Delta_{s}$ in
Fig.~\ref{sfig:output-photon-flux-strong}. As expected, for a
given mechanical driving, we can observe a resonance peak,
corresponding to the maximum value of the photon flux. This
behavior in the laboratory frame can directly reflect the behavior
of the excitation spectrum $\average{a_{s}^{\dag}a_{s}}_{\rm
    ss}\left(\Delta_{s}\right)$ in the squeezed frame in
Figs.~\ref{sfig:omega_m=omega_d}(a)
and~\ref{sfig:omega_m=2omega_s}(a). This is because the background
noise $\Phi_{\rm BGN}$ remains unchanged when the detuning is
changed, and the peak completely arises from the DCE in the
squeezed frame. Thus, the appearance of the peak of the output
flux spectrum $\Phi_{\rm out}\left(\Delta_{s}\right)$ can be
considered as an experimentally observable signature of the DCE.

\subsection{Signal-to-noise ratio and second-order correlation function at resonance}

\begin{figure}[h]
    \centering
    \includegraphics[width=8.5cm]{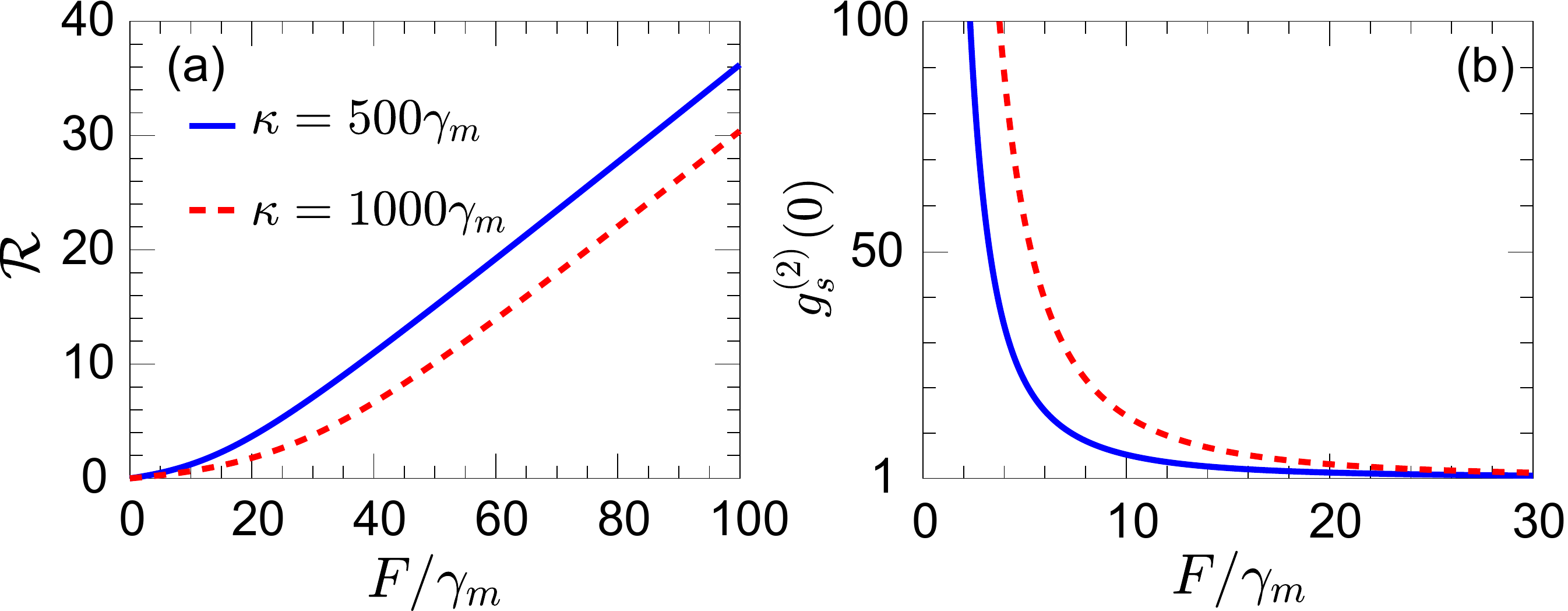}
    \caption{(a) Signal-to-noise ratio $\mathcal{R}$ and (b) correlation function $g_{s}^{\left(2\right)}\left(0\right)$ versus the mechanical driving $F$ for $\kappa=500\gamma_{m}$ (solid curves) and $1000\gamma_{m}$ (dashed curves). For both plots, we assumed that $\omega_{m}=\omega_{d}=2\omega_{s}$, $g_{0}=10\gamma_{m}$, and $\sinh^{2}\left(r\right)=0.5$.}\label{sfig:SNR_plus_g2}
\end{figure}

As mentioned before, there exists a background noise $\Phi_{\rm
BGN}$ in the flux $\Phi_{\rm out}$. Thus, we need to analyze the
ability of our proposal to resolve the DCE-induced signal from the
background noise. To quantitatively describe this ability, we
typically employ the signal-to-noise ratio defined in
Eq.~(\ref{eq:signal_to_noise_ratio}). Without loss of generality,
we focus on the ratio $\mathcal{R}$ at resonance
$\omega_{m}=\omega_{d}=2\omega_{s}$. Under this resonance
condition, the cubic equation satisfied by
$\average{a_{s}^{\dag}a_{s}}_{\rm ss}$ becomes
\begin{align}\label{seq:cubic-eq-at-resonance}
0=\;&g_{\rm DCE}^{4}\,x^3-g_{\rm DCE}^{2}\left(g_{\rm DCE}^{2}-\gamma_{0}^{2}\right)x^{2}\nonumber\\
&+\left[\frac{\gamma_{0}^{4}}{4}-g_{\rm
DCE}^{2}\left(\gamma_{0}^{2}+F^{2}\right)\right]x-\frac{\gamma_{0}^{4}}{4},
\end{align}
where $x=2\average{a_{s}^{\dag}a_{s}}+1$. Then,
$\average{a_{s}^{2}}_{\rm ss}$ and $\average{b}_{\rm ss}$ are
given by
\begin{align}
\average{a_{s}^{2}}_{\rm ss}&=-\frac{g_{\rm DCE}}{2g_{\rm DCE}^{2}\,x+\gamma_{0}^{2}}Fx,\\
\average{b}_{\rm ss}&=-\frac{i\kappa}{2g_{\rm
DCE}^{2}\,x+\gamma_{0}^{2}}\frac{F}{2}.
\end{align}
We plot the ratio $\mathcal{R}$ versus the driving $F$ in
Fig.~\ref{sfig:SNR_plus_g2}(a). We find that the signal-to-noise
ratio monotonically increases with the mechanical driving. This is
owing to the fact that an increase in the mechanical driving leads
to an increase in the number of DCE-induced photons, but at the
same time leaves the number of background-noise photons unchanged.

The equal-time second-order correlation function is defined in
Eq.~(\ref{eq:g2_correlation}). Similarly to the discussion of the
signal-to-noise ratio $\mathcal{R}$, we also only focus on the
$g_{s}^{\left(2\right)}\!\left(0\right)$ correlation at resonance
$\omega_{m}=\omega_{d}=2\omega_{s}$. In the semi-classical
treatment presented in this section, $\average{a_{s}^{\dag
2}a_{s}^{2}}_{\rm ss}$ can be approximated as
$\average{a_{s}^{\dag 2}a_{s}^{2}}_{\rm
ss}\approx|\average{a_{s}^{2}}_{\rm ss}|^{2}$, and as a result,
the $g_{s}^{\left(2\right)}\!\left(0\right)$ correlation is
reduced to
\begin{equation}
g_{s}^{\left(2\right)}\!\left(0\right)\approx\frac{|\average{a_{s}^{2}}_{\rm
ss}|^{2}}{\average{a_{s}^{\dag}a_{s}}^{2}_{\rm ss}},
\end{equation}
which is plotted as a function of the mechanical driving in
Fig.~\ref{sfig:SNR_plus_g2}(b). We find that
$g_{s}^{\left(2\right)}\!\left(0\right)$ starts with very large
values, and as the mechanical driving increases,
$g_{s}^{\left(2\right)}\!\left(0\right)$ then decreases
approaching 1. This behavior, as expected, suggests the phenomenon
of photon bunching, thus confirming the DCE.

\subsection{Analytical solutions in the limits $F\rightarrow0$ and $F\rightarrow\infty$}

In order to have a better analytical understanding, let us now
consider the limit of $F\rightarrow0$, and also the opposite limit
of $F\rightarrow\infty$, at resonance
$\omega_{m}=\omega_{d}=2\omega_{s}$.

For the $F\rightarrow0$ limit, we have
$\average{a_{s}^{\dag}a_{s}}_{\rm ss}\rightarrow0$, and thus,
$x^{n}\approx1+2n\average{a_{s}^{\dag}a_{s}}$ for
$n=0,1,2,\cdots$. Based on this, an approximate solution of the
cubic equation in Eq.~(\ref{seq:cubic-eq-at-resonance}) is found
to be
\begin{equation}
\average{a_{s}^{\dag}a_{s}}_{\rm ss}\approx\frac{2g_{\rm
DCE}^{2}}{\left(2g_{\rm DCE}^{2}+\gamma_{0}^{2}\right)^{2}}F^{2},
\quad {\rm when} \quad F\rightarrow0,
\end{equation}
which corresponds to
Eq.~(\ref{seq:steady-state-intracavity-photon-number}) for $n_{\rm
th}=0$. Analogously, we obtain
\begin{align}
\label{seq:DCE-induced-asas_02}
\average{a_{s}^{2}}_{\rm ss}&\approx-\frac{g_{\rm DCE}}{2g_{\rm DCE}^{2}+\gamma_{0}^{2}}F,\quad {\rm when} \quad F\rightarrow0,\\
\average{b}_{\rm ss}&\approx-\frac{i\kappa}{2g_{\rm
DCE}^{2}+\gamma_{0}^{2}}\frac{F}{2}, \quad {\rm when} \quad
F\rightarrow0.
\end{align}
Note that Eq.~(\ref{seq:DCE-induced-asas_02}) corresponds to
Eq.~(\ref{seq:DCE-induced-asas}). Therefore, according to
Eq.~(\ref{seq:steady-state-output}), we obtain a quadratic
increase in the ratio
\begin{equation}
\mathcal{R}=\frac{\Phi_{\rm DCE}}{\Phi_{\rm BGN}}\propto F,
\end{equation}
with large driving $F$, as shown in
Fig.~\ref{sfig:SNR_plus_g2}(a).

In the opposite limit of $F\rightarrow\infty$, we have
$x\rightarrow2\average{a_{s}^{\dag}a_{s}}_{\rm ss}$, and then
obtain
\begin{align}
\average{a_{s}^{\dag}a_{s}}_{\rm ss}&\approx \frac{F}{2g_{\rm DCE}}, \quad {\rm when} \quad F\rightarrow\infty,\\
\average{a_{s}^{2}}_{\rm ss}&\approx-\frac{F}{2g_{\rm DCE}}, \quad {\rm when} \quad F\rightarrow\infty,\\
\average{b}_{\rm ss}&\approx-\frac{i\kappa}{4g_{\rm DCE}}, \quad
{\rm when} \quad F\rightarrow\infty.
\end{align}
Consequently, the photon flux $\Phi_{\rm out}$ is given by
\begin{equation}
\Phi_{\rm DCE}=\frac{F}{2g_{\rm DCE}}\exp\left(2r\right), \quad
{\rm when} \quad F\rightarrow\infty.
\end{equation}
This indicates a linear increase in the ratio $\mathcal{R}$ with
the driving $F$, as shown in Fig.~\ref{sfig:SNR_plus_g2}(a).

For the $g_{s}^{\left(2\right)}\!\left(0\right)$ correlation in
the limit of $F\rightarrow0$, we find
\begin{equation}
g_{s}^{\left(2\right)}\!\left(0\right)\approx\frac{1}{2\average{a_{s}^{\dag}a_{s}}_{\rm
ss}}, \quad {\rm when} \quad F\rightarrow0,
\end{equation}
which is the same as
Eq.~(\ref{seq:second-order-correlation-weak}). This corresponds to
a large $g_{s}^{\left(2\right)}\!\left(0\right)$ as in
Fig.~\ref{sfig:SNR_plus_g2}(b) and, thus, to large photon
bunching.

Furthermore, in the opposite limit of $F\rightarrow\infty$, the
correlation function $g_{s}^{\left(2\right)}\!\left(0\right)$ is
approximately equal to $1$, i.e.,
\begin{equation}
g_{s}^{\left(2\right)}\!\left(0\right)\approx1, \quad {\rm when}
\quad F\rightarrow\infty,
\end{equation}
as shown Fig.~\ref{sfig:SNR_plus_g2}(b). This means that the DCE
radiation field is approximately in a coherent state.

\subsection{Stability analysis}

\begin{figure}[t]
    \centering
    \includegraphics[width=7.0cm]{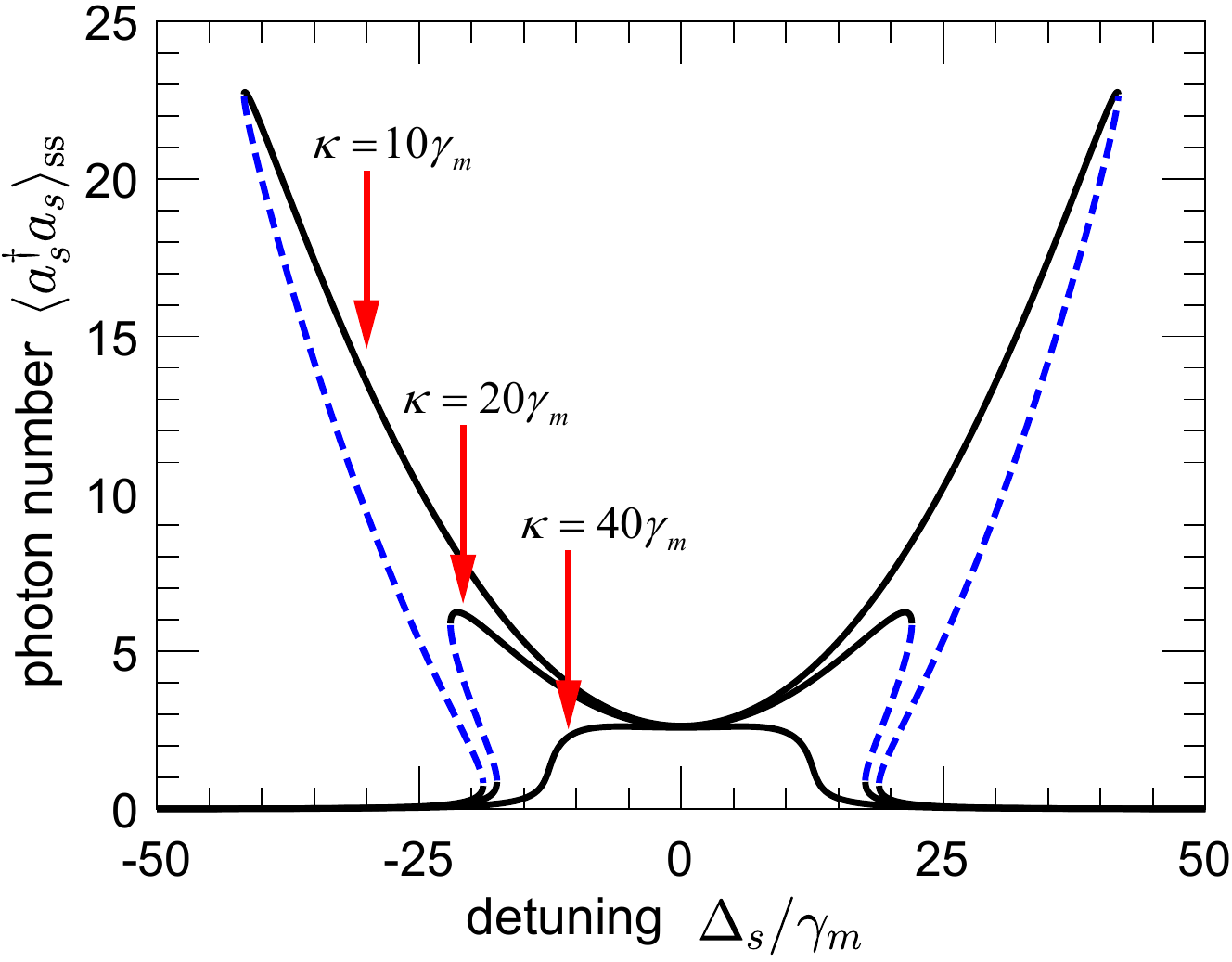}
    \caption{Steady-state squeezed cavity mode photon number
        $\average{a_{s}^{\dag}a_{s}}_{\rm ss}$ as a function of the
        detuning $\Delta_{s}$ for $\kappa=10\gamma_{m}$, $20\gamma_{m}$,
        and $40\gamma_{m}$. Dashed curves represent unstable solutions.
        Here, we assumed that $\omega_{m}=2\omega_{s}$,
        $g_{0}=10\gamma_{m}$, $F=50\gamma_{m}$, and
        $\sinh^{2}\left(r\right)=0.5$.}\label{sfig:multistability}
\end{figure}

We now turn to multistability effects of our system. As discussed
previously, in the semi-classical approximation, the system is
governed by a cubic function. However, a cubic function has three
solutions, and thus the system may exhibit multistability effects.
To analyze them, we need to perform steady-state
analysis~\cite{sarchi2008coherent}. Thus, we express the
quantities $\average{a_{s}^{\dag}a_{s}}$, $\average{a_{s}^{2}}$,
and $\average{b}$ as the sum of their steady-state values
($\average{a_{s}^{\dag}a_{s}}_{\rm ss}$, $\average{a_{s}^{2}}_{\rm
ss}$, $\average{b}_{\rm ss}$) and time-dependent small
perturbations [$\delta_{1}\left(t\right)$,
$\delta_{2}\left(t\right)$, $\delta_{3}\left(t\right)$], that is,
\begin{align}
\average{a_{s}^{\dag}a_{s}}&=\average{a_{s}^{\dag}a_{s}}_{\rm ss}+\delta_{1}\left(t\right),\\
\average{a_{s}^{2}}&=\average{a_{s}^{2}}_{\rm ss}+\delta_{2}\left(t\right),\\
\average{b}&=\average{b}_{\rm ss}+\delta_{3}\left(t\right).
\end{align}
Then, substituting these equations into
Eqs.~(\ref{seq:semiclassical-equation01}),
(\ref{seq:semiclassical-equation02}), and
(\ref{seq:semiclassical-equation03}) yields
\begin{align}
\label{seq:small-fluctuation01}
\frac{d}{dt}\delta_{1}\left(t\right)=\;&i2g_{\rm DCE}\big(\average{b}_{\rm ss}^{*}\delta_{2}+\average{a_{s}^{2}}_{\rm ss}\delta_{3}^{*}\nonumber\\
&-\average{b}_{\rm ss}\delta_{2}^{*}-\average{a_{s}^{2}}_{\rm ss}^{*}\delta_{3}\big)-\kappa\delta_{1},\\
\label{seq:small-fluctuation02}
\frac{d}{dt}\delta_{2}\left(t\right)=&-i2\Delta_{s}\delta_{2}-i2g_{\rm DCE}\left(2\average{a_{s}^{\dag}a_{s}}_{\rm ss}+1\right)\delta_{3}\nonumber\\
&-i4g_{\rm DCE}\delta_{1}\average{b}_{\rm ss}-\kappa\delta_{2},\\
\label{seq:small-fluctuation03}
\frac{d}{dt}\delta_{3}\left(t\right)=&-i\Delta_{m}\delta_{3}-ig_{\rm
DCE}\delta_{2}-\frac{\gamma_{m}}{2}\delta_{3}.
\end{align}
We further make the following replacements,
\begin{align}
\delta_{1}\left(t\right)&\mapsto \exp\left(-i\omega t\right)x_{1}+\exp\left(i\omega^{*}t\right)y_{1}^{*},\\
\delta_{2}\left(t\right)&\mapsto \exp\left(-i\omega t\right)x_{2}+\exp\left(i\omega^{*}t\right)y_{2}^{*},\\
\delta_{3}\left(t\right)&\mapsto \exp\left(-i\omega
t\right)x_{3}+\exp\left(i\omega^{*}t\right)y_{3}^{*},
\end{align}
where $x_{k}$ and $y_{k}$ ($k=1,2,3$) are time-independent complex
numbers, and $\omega$ denotes a complex frequency. Then, the
coupled equations~(\ref{seq:small-fluctuation01}),
(\ref{seq:small-fluctuation02}), and
(\ref{seq:small-fluctuation03}) can be rewritten as
\begin{equation}
M\Psi=\omega\Psi,
\end{equation}
where
\begin{equation}
\Psi=\left(x_{1},y_{1},x_{2},y_{2},x_{3},y_{3}\right)^{T},
\end{equation}
\begin{widetext}
\begin{align}
M=i\left(
\begin{array}{cccccc}
-\kappa & 0 & A^{*} & A & B^{*} & B\\
0& -\kappa & A^{*} & A & B^{*} & B\\
2A & 0 & -i2\Delta_{s}-\kappa  &   0   & C & 0\\
0 & 2A^{*} & 0  &  i2\Delta_{s}-\kappa   & 0 & C^{*}\\
0 & 0 & -ig_{\rm DCE} & 0 &-i\Delta_{m}-\gamma_{m}/2 & 0\\
0 & 0 & 0 & ig_{\rm DCE} & 0 & i\Delta_{m}-\gamma_{m}/2
\end{array}
\right),
\end{align}
\end{widetext}
where
\begin{align}
A&=-i2g_{\rm DCE}\average{b}_{\rm ss},\\
B&=i2g_{\rm DCE}\average{a_{s}^{2}}_{\rm ss},\\
C&=-i2g_{\rm DCE}\left(2\average{a_{s}^{\dag}a_{s}}_{\rm
ss}+1\right).
\end{align}
If all imaginary parts of the eigenvalues of the matrix $M$ are
negative, then the system is stable; otherwise the system is
unstable~\cite{kyriienko2014optomechanics}. According to this
criterion, we estimate the stability of our system. We find that
for the parameters used in the above discussion about the DCE, the
system does not exhibit multistability. Furthermore, when the
mechanical loss is close to the cavity loss, we find for
$\omega_{m}=2\omega_{s}$ that the system becomes multistable, as
shown in Fig.~\ref{sfig:multistability}. However, the requirement
that the mechanical loss is close to the cavity loss makes the
threshold
\begin{equation}
F_{\rm th}=g_{\rm DCE}+\kappa\gamma_{m}/4g_{\rm DCE}
\end{equation}
very low. For $F\gg F_{\rm th}$, the system behaves classically,
and quantum effects are negligible~\cite{wilson2010photon,
butera2019mechanical}.  For the parameters in
Fig.~\ref{sfig:multistability}, the value of $F_{\rm th}$ is
$\approx9\gamma_{m}$ (here, we set $\kappa=20\gamma_{m}$), which
is smaller than one fifth of the force $F=50\gamma_{m}$. As a
consequence, the system, when demonstrating such multistable
behaviors, has probably reached the classical regime, where the
DCE effect induced by the quantum fluctuations is negligible.
Therefore, in order to observe the DCE, it is better to avoid the
multistable regime of the system.

\section{Possible implementations with superconducting quantum circuits}

\label{sec:Possible-implementation-with-superconducting-circuits}
Our scheme to implement the DCE is based on a generic
optomechanical system, and at the same time, does not require an
ultra-high-frequency mechanical resonator and an ultrastrong
single-photon coupling between light and mechanical motion.
Therefore, we can expect that it can be implemented in various
physical systems. In this section, as an example, we discuss in
detail a possible implementation with superconducting circuits
and, in particular, we refer to the experimental superconducting
quantum circuit of Ref.~\cite{teufel2011sideband}, described by
the standard optomechanical coupling of the form
$a^{\dag}a\left(b+b^{\dag}\right)$.

\begin{figure*}[tbph]
    \centering
    \includegraphics[width=14.0cm]{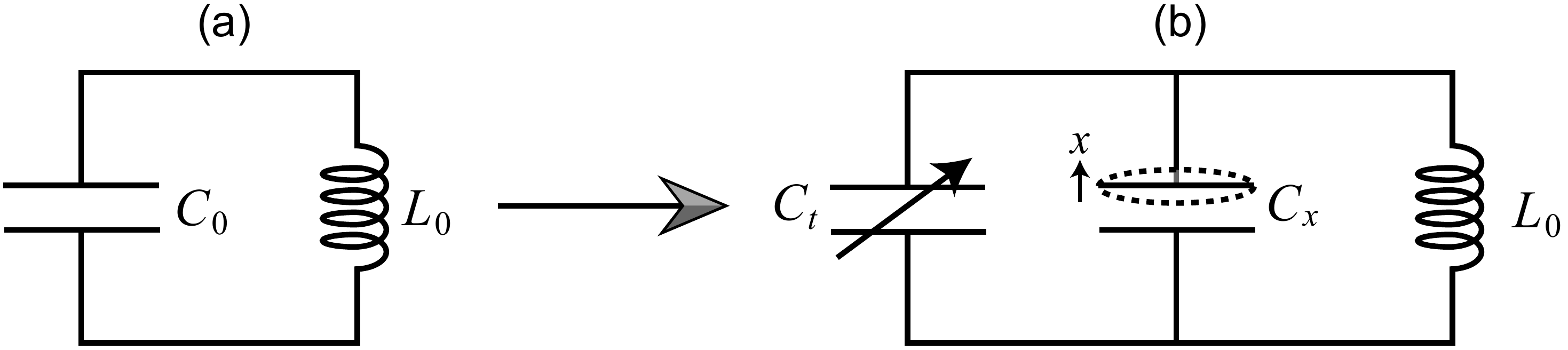}
    \caption{(a) A standard {\it LC} circuit consisting of a capacitor ($C_{0}$) and an inductor ($L_{0}$). This circuit behaves as a single-mode microwave cavity. (b) An {\it LC} circuit used to implement the DCE. Its capacitor ($C_{x}$) is modulated by the mechanical motion, e.g., of a micromechanical membrane, and this results in a standard optomechanical coupling between light and mechanics. Meanwhile, the use of the electrically tunable capacitor ($C_{t}$) can parametrically drive and squeeze the cavity mode.}\label{sfig-LC-circuits}
\end{figure*}

A standard {\it LC} circuit consists of a capacitor (e.g., with
capacitance $C_{0}$) and an inductor (e.g., with inductance
$L_{0}$), as shown in Fig.~\ref{sfig-LC-circuits}(a).  Its
Hamiltonian is expressed in terms of the capacitor charge $Q$ and
the inductor current $I$ as
\begin{equation}\label{seq:simple-LC-Hamiltonian}
H_{0}=\frac{\Phi^{2}}{2L_{0}}+\frac{1}{2}L_{0}\omega_{0}^{2}Q^{2},
\end{equation}
where $\Phi=L_{0}I$ is the magnetic flux through the inductor, and
$\omega_{0}=1/\sqrt{L_{0}C_{0}}$ is the fundamental frequency of
the circuit. After quantization, the charge $Q$ and the flux
$\Phi$ represent a pair of canonically conjugate variables, which
obey the commutation relation $\left[Q, \Phi\right]=i\hbar$. Upon
introducing a canonical transformation,
\begin{align}\label{seq:canonical-transformation}
Q=&\frac{1}{2}\sqrt{2\hbar\omega_{0}C_{0}}\left(a+a^{\dag}\right),
\nonumber\\
\Phi=&\frac{1}{2i}\sqrt{2\hbar\omega_{0}L_{0}}\left(a-a^{\dag}\right),
\end{align}
the Hamiltonian $H_{0}$ becomes
\begin{equation}
H_{0}=\hbar\omega_{0}a^{\dag}a.
\end{equation}
Here, we have subtracted the constant zero-point energy
$\hbar\omega_{0}/2$. Such an {\it LC} circuit thus behaves as a
single-mode microwave cavity, with $\omega_{0}$ being the cavity
frequency, and with $a$ ($a^{\dag}$) being the annihilation
(creation) operator of the cavity mode.

As demonstrated in Ref.~\cite{teufel2011sideband}, when the
capacitance $C_{0}$ in Fig.~\ref{sfig-LC-circuits}(a) is modulated
by the mechanical motion of a micromechanical membrane, the
mechanical motion can couple to the cavity mode. In this manner,
the capacitance $C_{0}$ becomes
\begin{equation}
C_{0}\mapsto C_{x}=\frac{C_{0}}{1+x/d},
\end{equation}
where $x$ is the displacement of the membrane, and $d$ is the
distance between the conductive plates of the capacitor. To
parametrically squeeze the cavity mode, we further add an
additional and electrically tunable capacitor into such an
experimental setup. The {\it LC} circuit is shown in
Fig.~\ref{sfig-LC-circuits}(b). Here, we assume the capacitance of
the additional capacitor to be
\begin{equation}
C_{t}=C_{0}+\Delta C\cos\left(\omega_{L}t\right),
\end{equation}
where $\omega_{L}$ is the modulation frequency, and $\Delta C\ll
C_{0}$. The total capacitance is thus given by $C_{\rm
total}=C_{t}+C_{x}$. Note that, in the absence of both mechanical
motion and cosine modulation, the total capacitance is equal to
$2C_{0}$, and as a result, the resonance frequency of the bare
{\it LC} cavity, shown  in Fig.~\ref{sfig-LC-circuits}(b), is
$\omega_{c}=\omega_{0}/\sqrt{2}$, rather than $=\omega_{0}$. When
both mechanical motion and cosine modulation are present, the
cavity frequency $\omega_{c}$ is modulated as
\begin{align}
\omega_{c}\mapsto\omega_{c}^{\prime}=&\frac{1}{\sqrt{L_{0}C_{\rm total}}}\nonumber\\
=&\frac{\omega_{0}}{\sqrt{1+\frac{\Delta
C}{C_{0}}\cos\left(\omega_{L}t\right)+\frac{1}{1+x/d}}}.
\end{align}
In the limit $\left\{\Delta C/C_{0}, x/d\right\}\ll1$, we can
expand $\omega_{c}^{\prime}$, up to first order, to have
\begin{equation}
\omega_{c}^{\prime}\approx\omega_{c}\left[1-\frac{\Delta
C}{4C_{0}}\cos\left(\omega_{L}t\right)+\frac{x}{4d}\right].
\end{equation}
The Hamiltonian describing the cavity mode of the {\it LC} circuit
in Fig.~\ref{sfig-LC-circuits}(b) is then given by
\begin{equation}
H_{c}=\frac{\Phi^{2}}{2L_{0}}+\frac{1}{2}L_{0}\omega_{c}^{2}\left[1-\frac{\Delta
C}{2C_{0}}\cos\left(\omega_{L}t\right)+\frac{x}{2d}\right]Q^{2}.
\end{equation}
Using the canonical transformation in
Eq.~(\ref{seq:canonical-transformation}), but with $\omega_{0}$
replaced by $\omega_{c}$, the Hamiltonian $H_{c}$ is reduced to
\begin{align}\label{seq:LC-cavity-mode-Hamiltonian}
H_{c}=\;&\hbar\omega_{c}a^{\dag}a-\hbar g_{0}a^{\dag}a\left(b+b^{\dag}\right)\nonumber\\
&+\frac{1}{2}\hbar\Omega\left[\exp\left(i\omega_{L}t
\right)a^{2}+\exp\left(-i\omega_{L}t\right)a^{\dag2}\right],
\end{align}
where $b$ ($b^{\dag}$) is the annihilation (creation) operator of
the mechanical mode, $g_{0}=-\omega_{c}x_{\rm zpf}/4d$ is the
single-photon optomechanical coupling, $x_{\rm zpf}$ is the
zero-point fluctuation of the mechanical resonator,
$\Omega=-\omega_{c}\Delta C/8C_{0}$ is the amplitude of the
two-photon driving, and $\omega_{L}$ is its frequency. Here, we
have made the rotating-wave approximation, and we have also
replaced
\begin{equation}
x\mapsto x_{\rm zpf}\left(b+b^{\dag}\right).
\end{equation}
After including the free Hamiltonian of the mechanical resonator,
the full Hamiltonian, in a rotating frame at $\omega_{L}/2$,
becomes ($\hbar=1$)
\begin{align}\label{seq:LC-Hamiltonian}
H=\;&\omega_{m}b^{\dag}b+\Delta a^{\dag}a\nonumber\\
&-g_{0}a^{\dag}a\left(b+b^{\dag}\right)+\frac{1}{2}\Omega\left(a^{2}+a^{\dag2}\right),
\end{align}
where $\omega_{m}$ is the frequency of the mechanical mode, and
$\Delta=\omega_{c}-\omega_{L}/2$. The Hamiltonian in
Eq.~(\ref{seq:LC-Hamiltonian}) is exactly the one applied by us in
this work.

A squeezed-vacuum reservoir coupled to the cavity mode can be
realized directly using the {\it LC} circuit in
Fig.~\ref{sfig-LC-circuits}(a), but the constant capacitance
$C_{0}$ needs to be replaced by a tunable capacitance $C_{t}$. By
following the same recipe as above, the corresponding Hamiltonian
is then given by
\begin{equation}
H_{r}=\Delta_{0}
a^{\dag}a+\frac{1}{2}\Omega_{0}\left(a^{2}+a^{\dag2}\right),
\end{equation}
where $\Delta_{0}=\omega_{0}-\omega_{L}/2$, and
$\Omega_{0}=-\omega_{0}\Delta C/8C_{0}$. The canonical
transformation used here is the same as given in
Eq.~(\ref{seq:canonical-transformation}). When the input field of
the cavity is in the vacuum, we can obtain a squeezed-vacuum field
at the output port, according to the input-output relation.

In addition to the {\it LC} circuit, the squeezed-vacuum reservoir
can also be generated by a Josephson parametric amplifier, as
experimentally demonstrated in
Refs.~\cite{murch2013reduction,toyli2016resonance}. In particular,
a squeezing bandwidth of up to $\sim 10$~MHz was reported in
Ref.~\cite{murch2013reduction}. This is sufficient to fulfil the
large-bandwidth requirement of the reservoir.


%

\end{document}